\RequirePackage{fix-cm}
\documentclass[smallextended]{svjour3}       
\smartqed  
\usepackage{graphicx}

\usepackage{cite}
\usepackage{hyperref}
\usepackage{adjustbox}
\usepackage{amssymb}
\usepackage{subfigure}
\usepackage{caption, verbatim}
\usepackage{enumitem}
\usepackage{graphicx}
\usepackage{amssymb}
\usepackage{multirow}
\usepackage{amsmath,mathtools}
\usepackage{longtable}
\usepackage{lscape}
\usepackage{tablefootnote}
\usepackage[table, dvipsnames]{xcolor}
\begin{document}

\title{Origins of ECG  and  Evolution of Automated DSP Techniques: A Review
}


\author{Neha Arora         \and
        Biswajit Mishra 
}



\date{Received: date / Accepted: date}

\maketitle

\begin{abstract}
 Over the years researchers have studied the evolution of Electrocardiogram (ECG) and the complex classification of cardiovascular diseases. This review focuses on the  evolution of the ECG, and covers the  most recent signal processing schemes with milestones over last 150 years in a systematic manner. Development phases of ECG, ECG leads, portable ECG monitors, Signal Processing Schemes and the Complex Transformations are discussed.   
 It also provides recommendations for the inclusion of certain important points based on the review.
\keywords{ECG Evolution \and Einthoven  \and Signal Processing \and Databases \and Automatic Classification}

\end{abstract}

\section{Introduction}
\label{intro}
In the year 1902, Dutch scientist  Willem Einthoven invented the String Galvanometer to measure the electrical cardiac activity that has become one of the most significant contributions of the century. This invention of ElectroCardioGram (ECG) or EleKtrocardioGram (EKG) revolutionizes the diagnoses of cardiovascular anomalies.  The history behind the subject covers similar developments occurring with oscillographs and electrometers. Therefore it is of interest to researchers to know the developments of ECG and its origin that led to the global acceptance of one of the most influential research of the century.

The invention of ECG spurred the research that was broadly divided into following schema:
\begin{itemize}
	\item Obtaining the ECG features for detection of cardiac anomalies.
	\item Number of leads or electrodes to acquire a complete 3-D view of the heart.
	\item  Portability of such systems to evaluate its uses in hospital and home settings.
\end{itemize}

The ECG signal's evolution is well studied and reported in \cite{Burnett, picture, stagesstg1}. In \cite{Burnett}, the authors discuss ECG background and the development of physiological instrumentation, oscillographs, String Galvanometer and the efforts by Cambridge Scientific Instruments (CSI) company to make ECG monitors practically usable in hospital settings. In \cite{picture}, the authors discuss Thomas Lewis's role for his efforts to make  ECG globally accepted, development of the String Galvanometer, Cambridge electrocardiograph machine, including the changes in the design for medical use. In \cite{stagesstg1} authors discuss the interesting history of ECG origin, electrophysiology practice in $19^{th}$ century, the contribution of  Willem Einthoven, String Galvanometer and early definitions of cardiac arrhythmias. It also discusses the American observations of the ECG, the role of Thomas Lewis and the development of Electrocardiography, ECG and Myocardial Infarction (MI), precordial leads, and augmented limb leads, Vectorcardiogram in clinical physiology and the challenges of Electrocardiography.  While these reviews focus on the historical development of the ECG, a recent review  \cite{reviewrecent}  focuses on various ECG signal processing research development that has occurred during 2000-2020. In \cite{reviewmi}, authors discuss several automatic detection methods for Myocardial Infarctions in their review.

In this review, we are focusing on the developments in the cardiac health monitoring schemes starting from ECG origin to the most recent ECG signal processing techniques. We have started with the work presented by Willem Einthoven and studied the history behind his work to cover the roadmap of ECG development. For, this we have studied the papers presented by William Einthoven and the pioneer researches in this field.
 Further, we have selected the most cited papers from Journals and conferences to follow the timeline of biomedical engineering development for cardiac health monitoring schemes.  During the years 2000-2020, an abundance of papers can be found in this stream related to front end development for ECG signal acquisition, ECG sensors, ECG electrodes, automatic signal processing with various compression techniques, etc. In this review, we have mainly considered the work related to the automatic signal processing domain because to the best of authors' knowledge, a broad review in this domain is not presented in the literature that covers the history of automatic signal processing development schemes.  We have not included the commercial systems available for cardiac health monitoring in the market in this review. As the main focus of the work is to determine recent trends of automatic signal processing techniques. Additionaly, the hardware of wearable devices is not the focus of the review. Interested readers can find the wearable devices review in \cite{Bansal}.
 
This review discusses the origin of ECG, further developments in the field and the automated ECG signal processing techniques available in the literature till date. The paper also discusses the limitations of existing machine learning and neural networks based classifiers based processing techniques. The tradeoff between the complexity of signal processing techniques, hardware and software system requirements for real-time systems is also discussed.  The shortcomings of utilizing the standard databases, in terms of different demographics, changing definitions of diseases over the period of time are also discussed in this review. 

\section{Origin of Electrocardiogram:}
\label{sec:1}
Advances in oscillographs were significant for the development of ECG as they provided the varying means for recording the alternating voltage.
The first oscillograph that enabled the recording of electrical variations by Blondel in 1893 is assumed to be the first predecessor of ECG \cite{blondel}. The electromechanical oscilloscope consisted of a moving part to provide oscillations for the detection of electrical current passing through it. He provided three probable solutions for the recording of electrical variations. The first approach was based on the moving magnet principle, the second on the moving coil and the last one was to adapt the telephones for the recording purpose. 
He chose the moving magnet approach to record the electrical variations till this time; moving coil galvanometers were exceptional ones and not being used for recording the electrical signal variations. 

\begin{figure}
	\centering
	\includegraphics[width=0.5\linewidth]{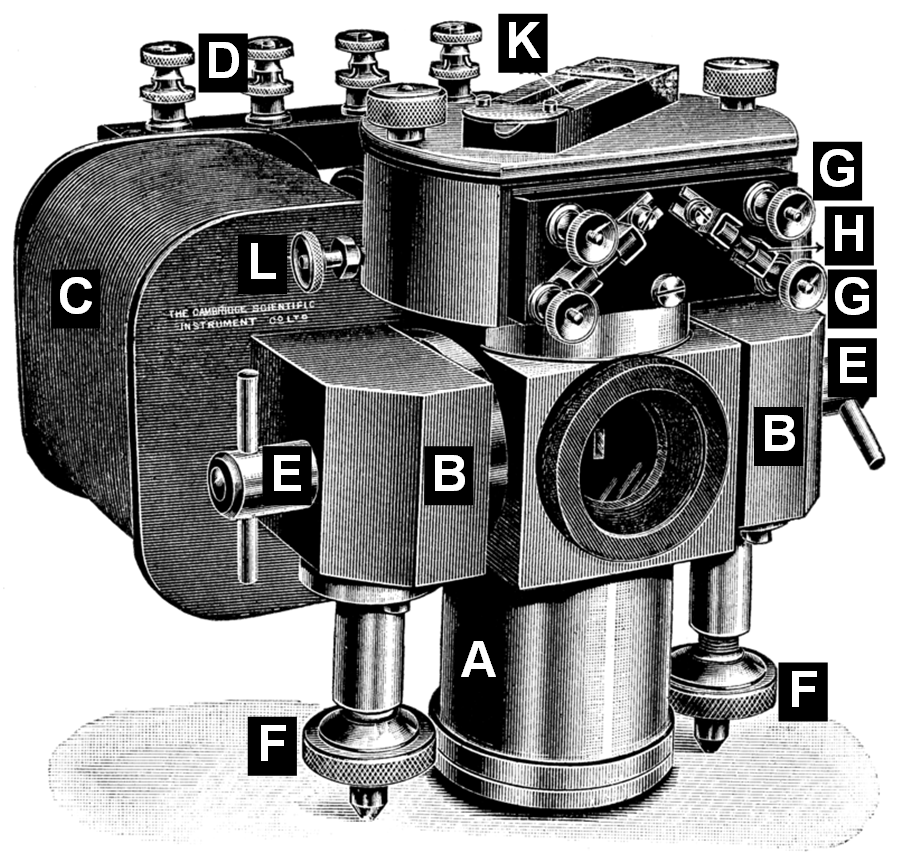}
	\caption{Dudells Oscillograph  (Picture Credit: \cite{Dudell})
		A is the brass oil bath in which two vibrators are fixed
		B, core of electromagnet which is excited by two coils
		C, one of the two coils
		D, two pairs of terminals for connecting the two coils
		E, bolts that hold the oil bath in position between the poles of the magnet
		F, two of three leveling screens (one is hidden behind the oil bath)
		G, terminals of one vibrator
		H, fuse
		K, thermometer with bulb in center of oil bath}
	\label{fig:dudell}
\end{figure}

The moving coil principle provided by the Blondel was the second generation of ECG. Dudell, in 1897 \cite{Dudell} replaced the conventional moving magnets with Phosphor Bronze Strips that utilizes the moving coil principle along with a mounted mirror that reflected a beam of light. The reflected beam fell on the photographic plate and provided a magnified recording of the movement of phosphor bronze strips. Dudell's oscillograph is shown in Fig. \ref{fig:dudell}.

In 1897, Clement Ader developed a Galvanometer \cite{Ader}, this significant invention aimed to boost the telegraphic transmission speed on long cables. The principle of transmission was based on very fine metal wires with $20 \ \mu m$ diameter, vibrating between the poles of large magnets. This galvanometer by Ader is perceived to be the first string galvanometer. When Willem Einthoven started experiments on recording the electrical activity of the heart in 1902, by his String Galvanometer, he was unaware that a similar system had already been developed by Clement Ader. Notably, the principle of operation for both the galvanometers was the same: a String was employed to record the electrical variation between large poles of magnets.  Einthoven's experiment was successful in its own right as later it was observed that the sensitivity of Ader's Galvanometer was lower compared to Einthoven's String Galvanometer. And this would not have been adequate for recording the physiological signals from the human body during the experiments. When Einthoven learned about Ader's String Galvanometer, he credited Ader and the researchers involved in these researches in one of his early papers \cite{raju}.

\begin{figure*}[htb!]
	\centering
	\includegraphics[scale=0.22]{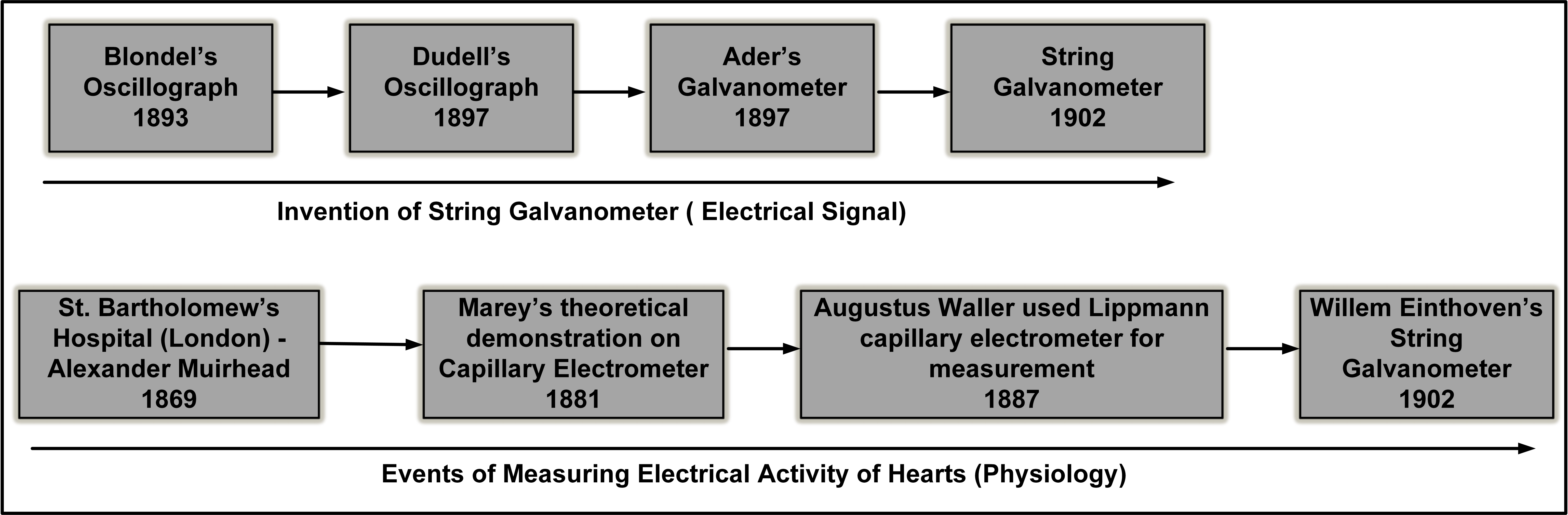}
	\caption{Precursor's of ECG and Events of Measuring Electrical Activity of Heart}
	\label{f1}
\end{figure*}

By now, several groups were already working on recording the electrical activity of the heart both from the signal and physiological point of view. Fig. \ref{f1} shows a brief overview of such events during the late 1800s to 1900.

\begin{figure}
	\centering
	\includegraphics[width=0.7\linewidth]{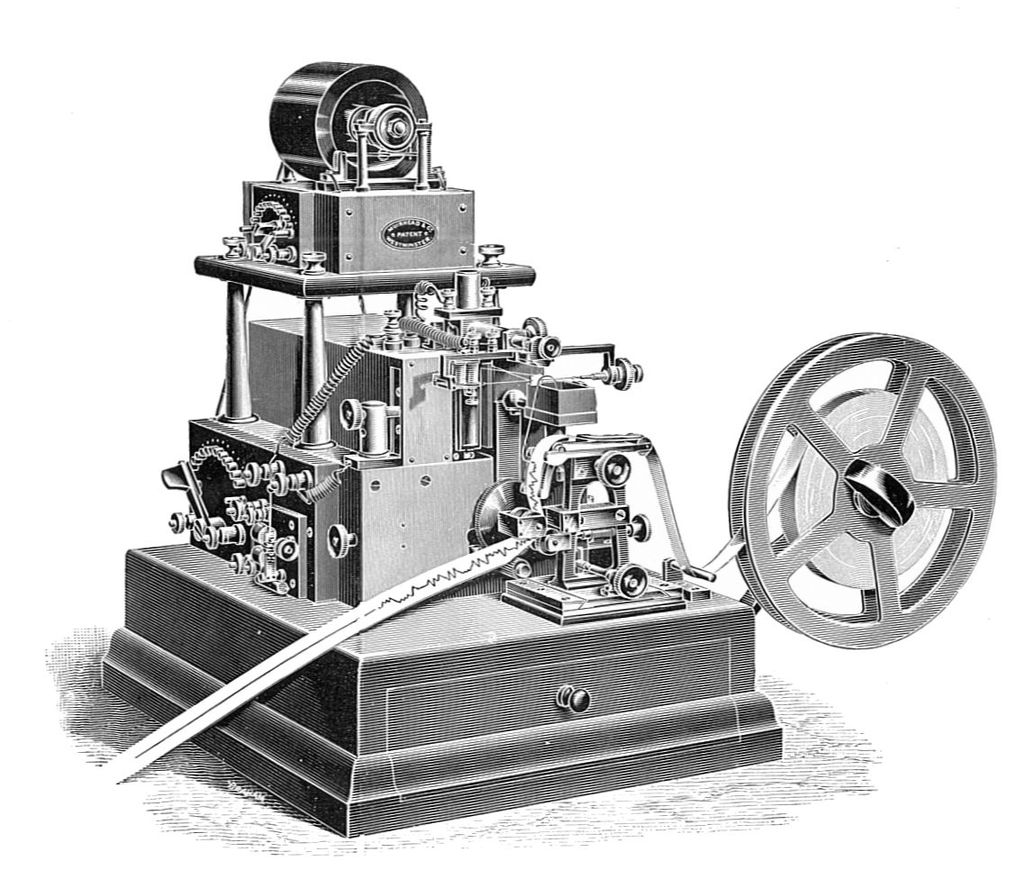}
	\caption{Alexander Muirhead ECG Research Instrument}
	\label{am}
\end{figure}

The first known successful event of recording the electrical activity of the heart was performed by Alexander Muirhead between  1869 - 1870 at St. Bartholomew’s Hospital in London using the Thomson Siphon Recorder \cite{thomson} as shown in Fig. \ref{am}. It was developed by William Thomson, a Telegraph Engineer. Muirhead recorded the ECG signal only once and after that, he never followed the research on physiological signals. 

Much before the Galvanometer experiments, Marey demonstrated theoretically in 1861 that it is possible to use a Capillary Electrometer to measure the electrical activity of the human heart. It was never investigated practically on humans, but the electrical activity of the heart of the frog was demonstrated using the Capillary Electrometer \cite{Marey1, Marey2}. The electrometer was invented by Gabriel Lipmann in 1872 while working in G. R. Kirchoff's laboratory in Berlin. It was used to detect small potential differences applied to thick end filled with mercury and thin end with the sulphuric acid solution (see Fig.\ref{fig:1}a ). 

\begingroup
\centering
\begin{figure}[htbp]
	\centering
	\subfigure[Capilary Electrometer]{\includegraphics[width=0.4\textwidth]{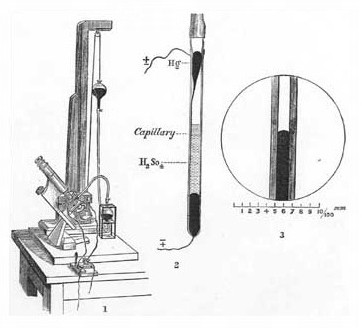}}\label{fig:1a}
	\subfigure[ECG Wave obtained by A.D. Waller Time Scale at Top, Pulse tracing in Centre and Electrocardiogram at Bottom] {\includegraphics[width=0.4\textwidth]{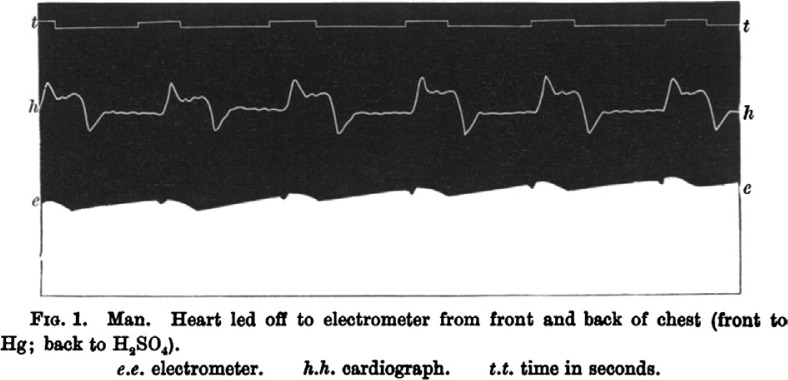}}\label{fig:1b}
	\subfigure[Experiment Demonstration to Royal Society by Waller's Pet Dog Jimmie (Illustrated London news May $22^{nd} \ May \ 1909$) (Picture Credit \cite{royal})]{\includegraphics[width=0.4\textwidth]{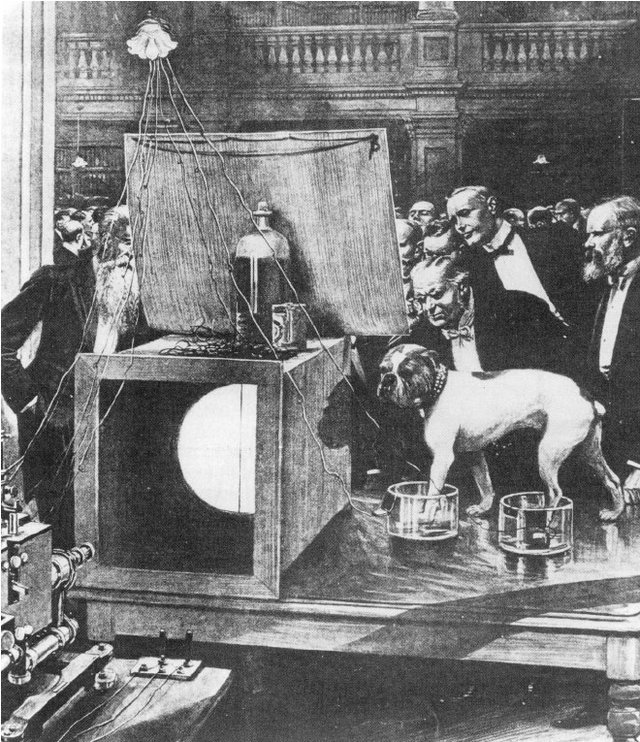}}\label{fig:1c}
	\caption{ Waller's Experimentation for ECG Tracing} \label{fig:1}
\end{figure}

\endgroup

Around the same period, Augustus D. Waller was using Marey's technique and applied it to exposed hearts of mammals. This led to the successful event of recording the electrical activity of the heart in 1887  using Lipmann Capillary Electrometer \cite{Waller}.
The experimental setup at Royal Society and the ECG wave obtained by Waller is shown in Fig. \ref{fig:1}. This event was widely publicized in local London Newspapers. Interestingly after this, he was investigated and tried under the `Animal Cruelty Act' for his experiments involving his pet dog, Jimmie.
Also, Waller himself was not convinced that it could be used widely in the biomedical domain. He went on further and stated,  ``\textit{I do not imagine that electrocardiography is likely to find any very extensive use in the hospital. . . It can at most be of rare and occasional use to afford a record of some rare anomaly of cardiac action.}"  Waller was not the first to use the term ``Electrocardiogram," and could not perhaps foresee this as ``the future" in Medical Technology and this is one of the reasons that his contributions are not acknowledged widely.

\subsection{  Willem Einthoven's String Galvanometer} 
By this time, the most notable research findings were coming out from Willem Einthoven’s String Galvanometer experiments \cite{Einth1} that measured cardiac potentials in 1902.  His experiments demonstrated that String Galvanometer was easier to use, free from damping, and more sensitive than the Capillary Electrometer. The String Galvanometer is shown in Fig. \ref{csistg}. His contributions revolutionized the biomedical cardiac signal acquisition forever. Einthoven was rightly entitled as the ``Father of Modern Electrocardiography". 

\begingroup
\centering
\begin{figure}[htbp]
	\centering
	\subfigure[String Galvanometer Picture Credit \cite{Einth1}  ]{\includegraphics[width=0.3\textwidth]{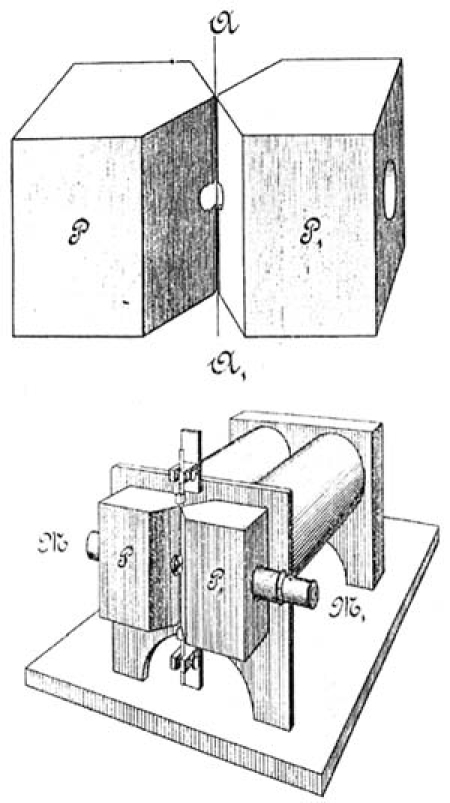}}\label{stg1}
	\subfigure[String Galvanometer Designed by Cambrige Scientific Instrument Limited Picture Credit:\cite{picture}] {\includegraphics[width=0.6\textwidth]{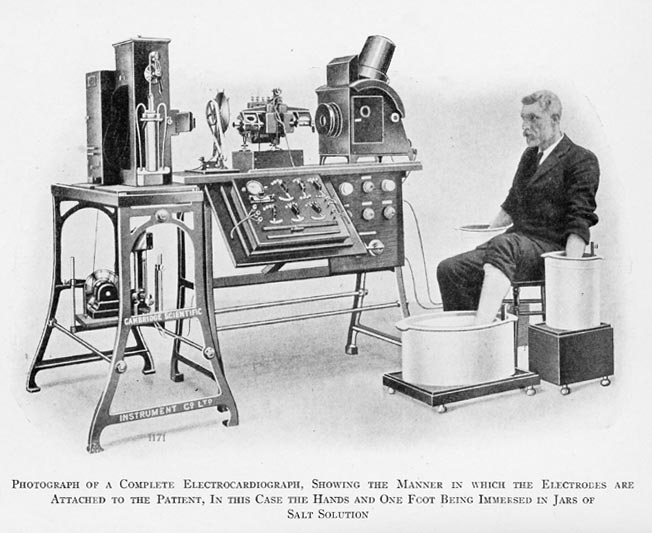}}\label{stg2}
	
	\caption{ String Galvanometer} 
	\label{csistg}
\end{figure}

\endgroup

The first deflections of the cardiac activity were named A, B, C, D by Einthoven \cite{Einth} shown in Fig. \ref{PQRST}a. A mathematically corrected version of deflections named as P, Q, R, S, T was superimposed on the former ones shown in Fig.\ref{PQRST}a  \cite{Einth0}. The naming conventions P, Q, R, S and T are still used to represent an ECG signal as shown in Fig. \ref{PQRST}b. The reason for changing the cardiac deflection's name from A, B, C, D to P, Q, R, S and T is still unclear but most likely it may have been done to include the successive points in the ECG tracings \cite{naming}. 

\begingroup
\centering
\begin{figure}[htbp]
	\centering
	\subfigure[P, Q, R, S and T deflections superimposed on previously known deflections A, B, C and D in the ECG Tracing By Einthoven ]{\includegraphics[width=0.45\textwidth]{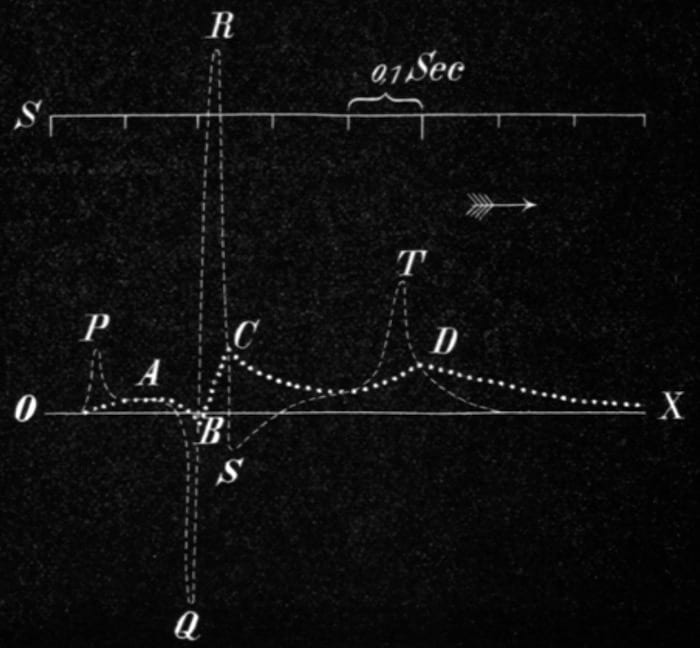}}\label{PQRST1}
	\subfigure[First ECG Tracing] {\includegraphics[width=0.45\textwidth]{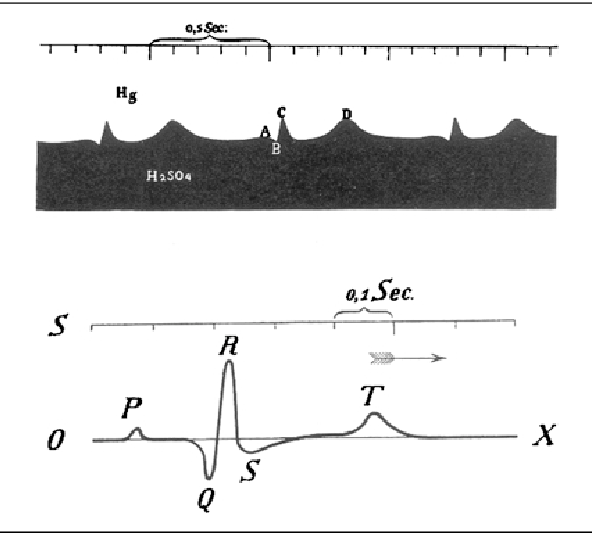}}\label{PQRST2}
	
	\caption{ First ECG Tracings by Einthoven (Picture Credit \cite{Einth1})} 
	\label {PQRST}
\end{figure}
\endgroup

According to  Einthoven, in \cite{Einth1}, the specifications for the String Galvanometer (Fig. \ref{csistg}a) is described as  ``\textit{The String Galvanometer is essentially composed of a thin silver‑coated quartz filament (about $3 \mu m$ thick), which is attached like a string in a strong magnetic field. When an electric current is conducted through this quartz filament, the filament reveals a movement that can be observed and photographed using considerable magnification, this movement is similar to the movement of the capillary electrometer. It is possible to regulate the sensitivity of the galvanometer very accurately within broad limits by tightening or loosening the string}''.

The original apparatus weighed around $600 \ lbs$ and required approximately two rooms for placing it. Provision for cooling the electromagnets was provided with continuous water flow. Saline water in buckets was used as electrodes on the left leg, left arm and right arm locations shown in Fig. \ref{csistg}b.  

\begin{figure}[htb!]
	\centering
	\includegraphics[scale=0.25]{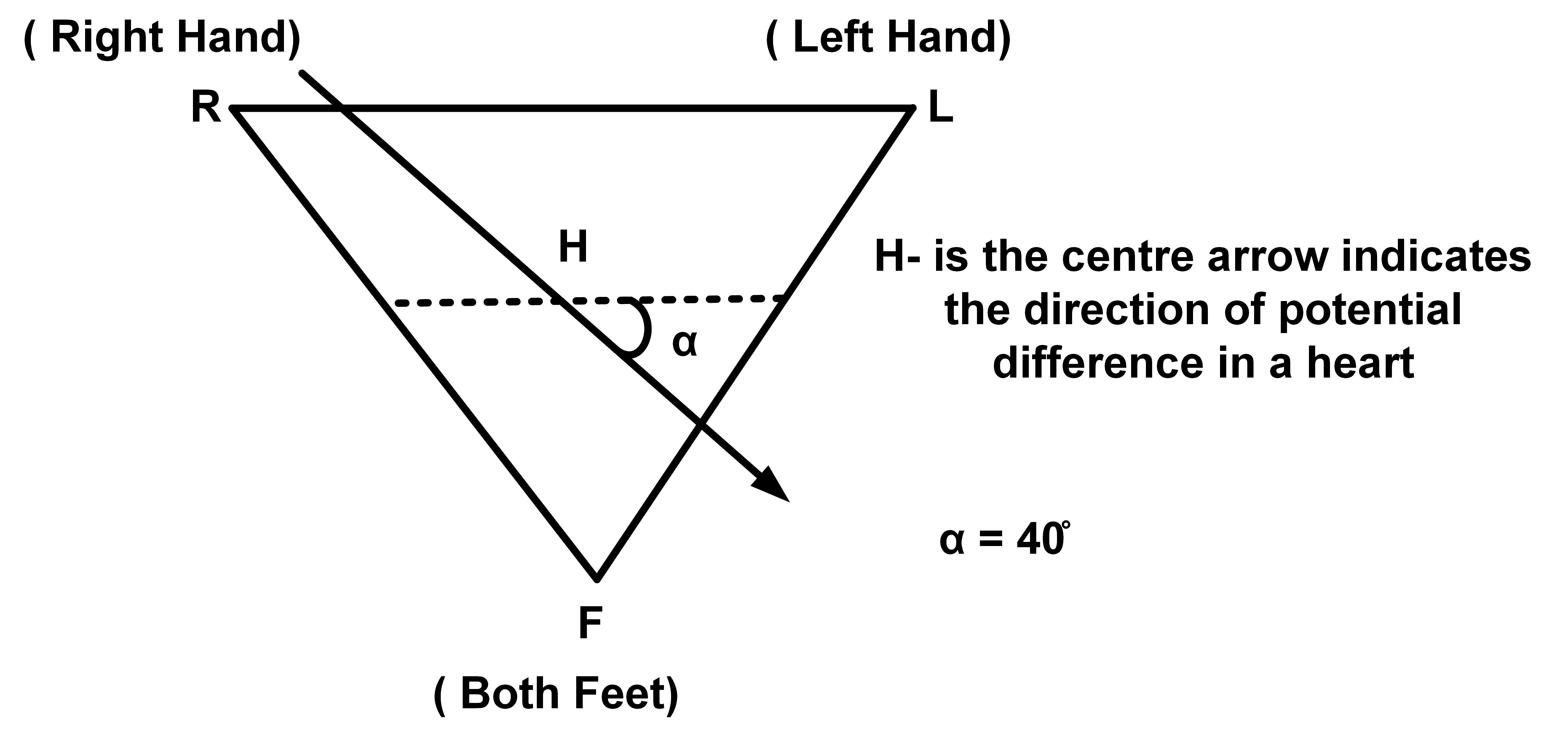}
	\caption{ Einthoven's Triangle}
	\label{et}
\end{figure}

Einthoven demonstrated significant differences in normal and abnormal ECG waveforms in 1906 \cite{Einth2} and 1908\cite{Einth3}. In the experiments to follow, in 1912  \cite{Einth4}, Einthoven investigated and found out that the heart creates a potential difference at different locations and the magnitude and direction of the current changes at different locations of the heart can be represented by the Einthoven Triangle is shown in Fig. \ref{et}.

He demonstrated that Lead I had advantages for judging the T waves, in Lead II peaks were usually larger and Lead III  was most suited for the diagnosis of  Ventricular hypertrophy \footnote{Ventricular hypertrophy is thickening of the walls of a ventricle (lower chamber) of the heart} of left and right ventricles.
He also observed a linear relationship between the three leads that yielded $ Lead \ II-Lead \ I= Lead \ III $ to obtain any lead by combining the other two leads.

\begin{figure}[htb!]
	\centering
	\includegraphics[scale=0.22]{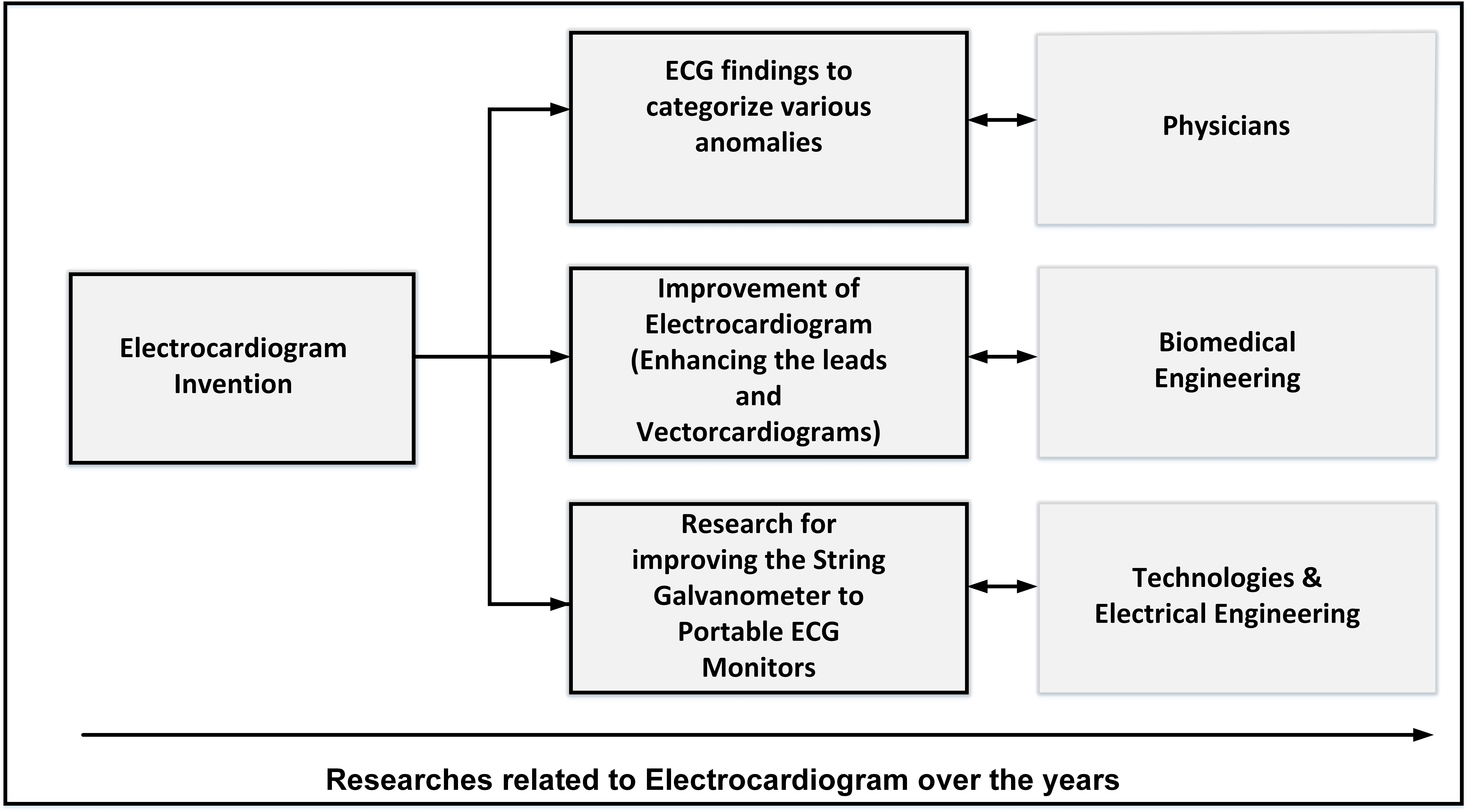}
	\caption{ EKG Evolution over the years}
	\label{f2}
\end{figure}

By then, Europe had accepted electrocardiograms and the rest of the world followed. 
The ECG research mainly evolved into three groups ( see Fig. \ref{f2}): those that were working to categorize the ECG signals for cardiac conditions: mainly consisted of Physicians; those working on optimizing electrodes and leads, and the third on optimizing the size of string galvanometer and designing portable bedside monitors, mostly technological experts.

\subsection{ Improving of String Galvanometer to Portable ECG Monitors }
The need to extend the ECG device to bedside monitors was a deriving factor for Einthoven. 
In 1903, he approached Cambridge Scientific Instruments Company Limited (CSI) to reduce the String Galvanometer's size to a more practicable form.  CSI became interested in the idea and produced their first String Galvanometer around 1905 (Fig. \ref{csistg}b) with 10\% royalty fees going to the inventor. The first instrument was sold to MacDonald's laboratory in Sheffield in 1905, the second to J.C. Bose at Presidency College, Calcutta in 1906 and the third to Keith Lucas in Cambridge in 1907. Later, Dudell made some modifications to the original design to reduce the instrument's size, which resulted in a reduced royalty to Einthoven. Edward Schafer of the University of Edinburgh bought the advanced version of the String Galvanometer and was the first to buy the string galvanometer for clinical use in 1908. Subsequently, after world war- I, this ECG machine evolved meant to be placed by the bedside. After a few modifications to the design, Harold Segall designed the instrument \cite{Burnett} that could be carried in two wooden cases,  each weighing around 50 lbs. This was the beginning of portable ECG monitors. 

The use of vacuum tubes for a reduced form factor and to amplify the electrocardiogram instead of the mechanical amplification by the string galvanometer is said to be used by 1925, reported by both Fye \cite{stagesstg1} and  Ernestine and Levine \cite{stagesstg2}. Subsequent advances in electronic components resulted in the first portable ECG machine by 1928, that was powered by a vehicle battery. This was until the invention of increasingly smaller transistor electronics \cite{stagesstg1}.  More recently, microchips allowed for developing the 12 lead ECG that we are familiar with today. Around 1935, Sunborn Company designed the ECG machine that could be kept in a wooden box weighing around 25 lbs \cite{Burnett}. The invention of transistors by the 1960s made the ECG machines portable for use in hospitals and the invention of Holter \footnote{Named on the scientist Norman Jeff Holter who developed it.} recorders in 1961 \cite{Holters} paved these for usage in out of hospital settings. 

\subsection{ ECG Signal (P Wave, QRS Complex, T wave, J point) : }

\begin{figure}[htb!]
	\centering
	\includegraphics[scale=0.22]{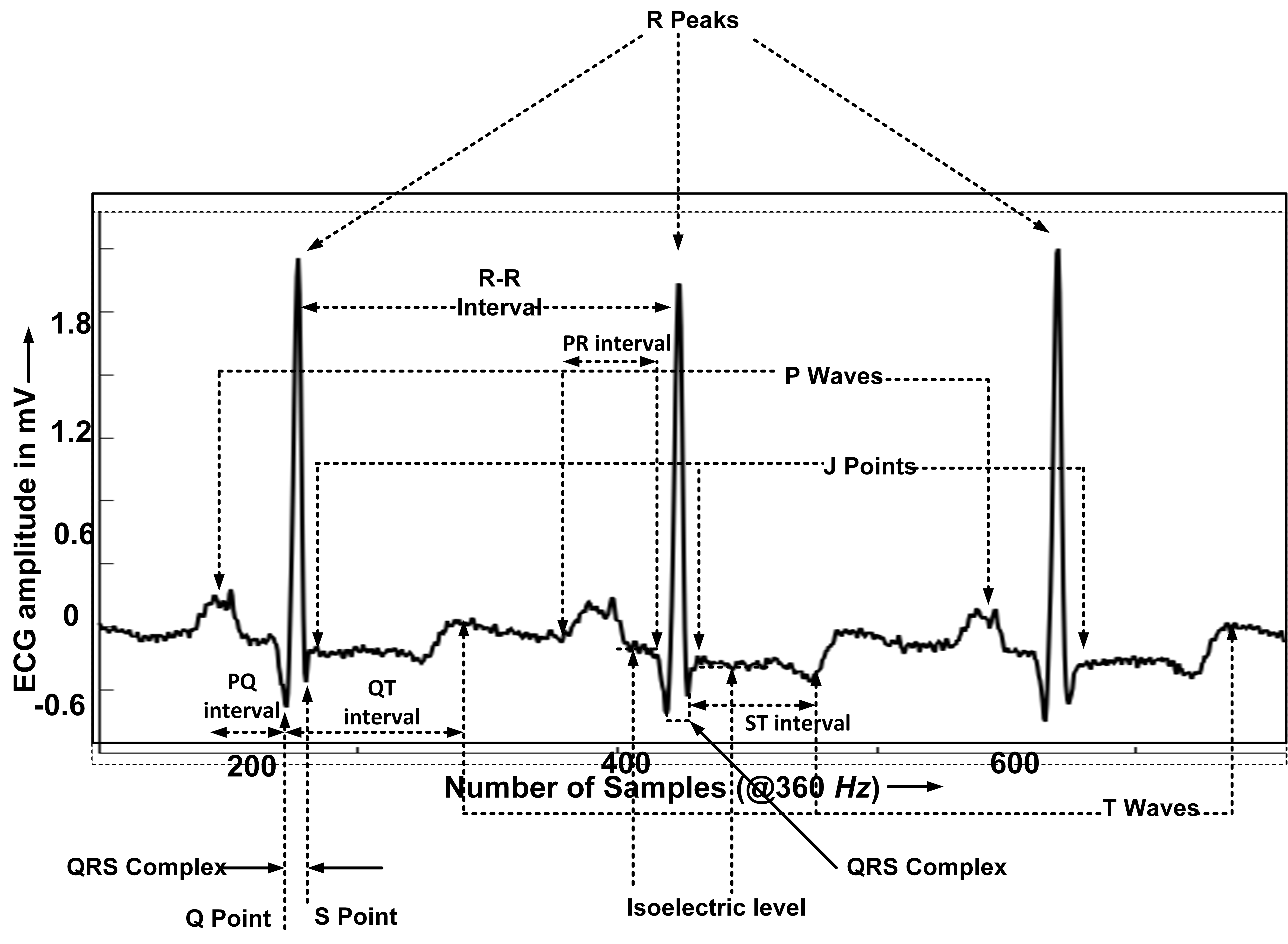}
	\caption{ P wave, QRS Complex, T wave, J point of an ECG signal }
	\label{pqrst}
\end{figure}
Following Einthoven's invention, scientists became interested in this emerging field and started categorizing the signals based on their interpretations. Fig. \ref{pqrst} shows the P wave, QRS complex, T wave, J point and Baseline level of the signal.

\begin{figure}[htb!]
	\centering
	\includegraphics[scale=0.60]{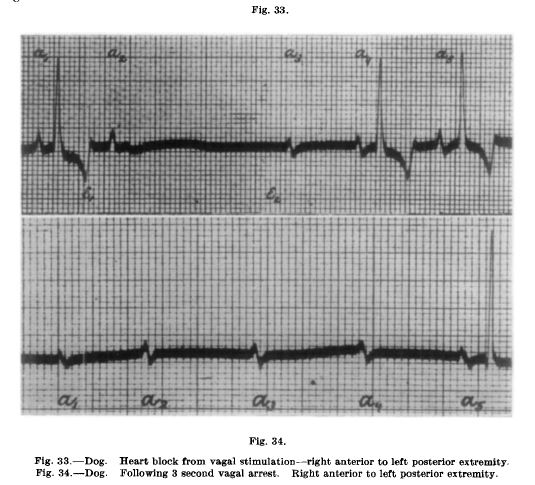}
	\caption{ECG Tracings of Experimentally Induced Heart Block in a Dog (Picture Credit \cite{Einthtele}) }
	\label{dog}
\end{figure}

Einthoven in 1906 categorized normal and abnormal ECGs that was translated by Cardiologist Henry Blackburn \cite{Einthtele}. He discussed the first electrocardiographic tracings of atrial fibrillation \footnote{ Cardiac anomaly where R-R interval is abnormal and P wave is missing at instances.},  premature ventricular contractions \footnote{ Abnormal heartbeat where contractions begin in the ventricles, instead of Sinoatrial node of heart}, ventricular bigeminy \footnote{Arrhythmia where there is a pattern of irregular heartbeat and regular heartbeat occurrence.}, atrial flutter \footnote{Type of arrhythmia, where the heart's upper chambers (atria) beat too quickly.}. The beginning of ECG related research also demonstrated in an experimental setup that induced heart block in a dog, as shown in Fig.\ref{dog}. 

\begin{figure}[htb!]
	\centering
	\includegraphics[scale=0.25]{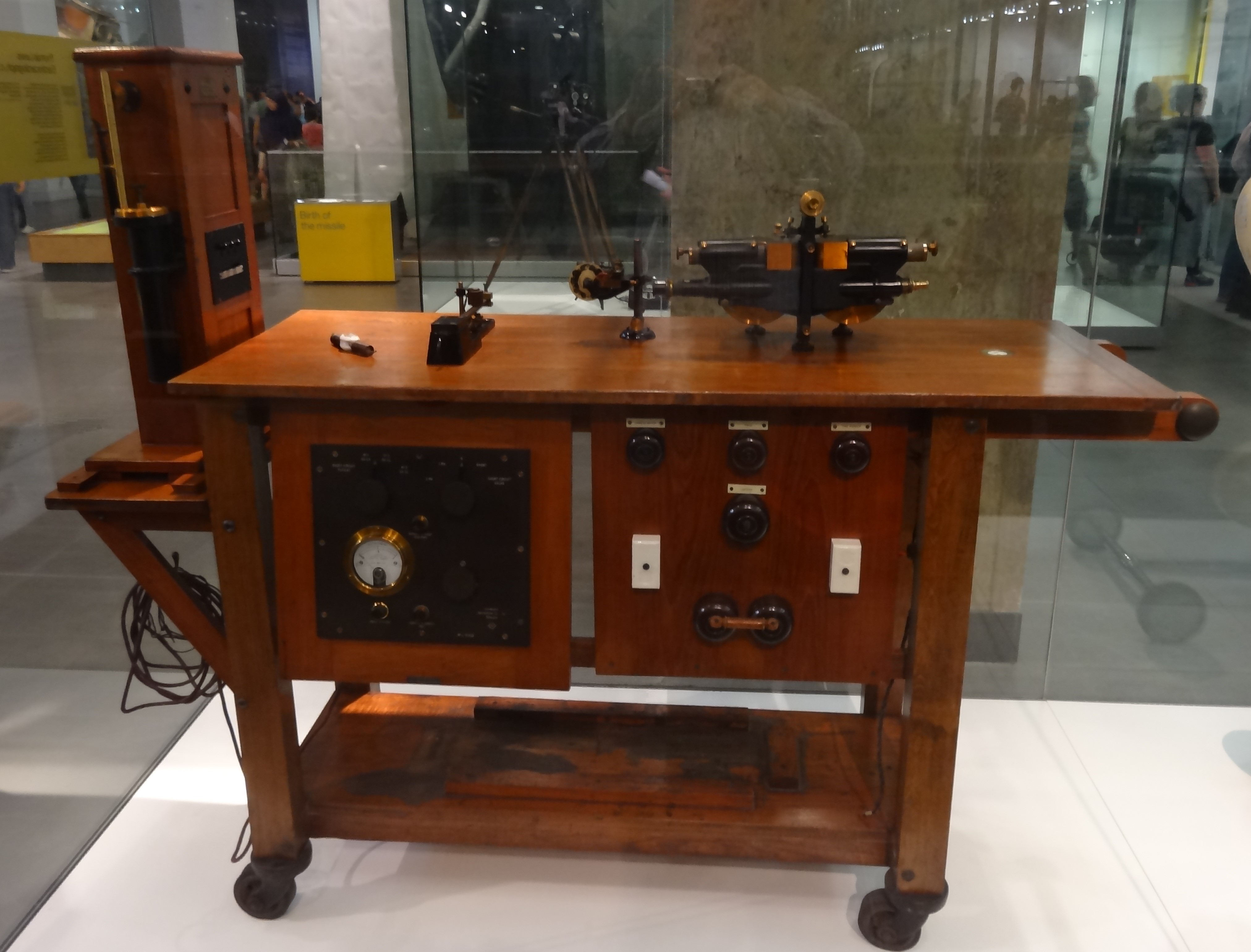}
	\caption{ECG Machine of Thomas Lewis for checking the patients in the year 1930. Picture from  Medical exhibits in the Science Museum, London (Picture Credit: Jonathan Cardy)}
	\label{tl3}
\end{figure}

By then, Thomas Lewis, a physician, was convinced about the significance of Einthoven's contribution to determine various heart anomalies.  Independently, he concluded that atrial fibrillation is a common cause of arrhythmia and termed as a ``clinical condition'' \cite{Lewis}. Fig. \ref{tl} shows the ECG signal of Atrial Fibrillation case and ECG waves of mother and fetal studied by Lewis \cite{picture}. Six major categories of anomalies were also coined by Lewis, namely: sinus arrhythmia, heart block, premature contractions, proximal tachycardia, auricular fibrillation and alteration of the pulse. The ECG machine used by Lewis during 1930 to diagnose the patients is shown in  Fig. \ref{tl3}. In addition he also explained the terms such as sino auricular node, pacemaker, premature contractions, proximal tachycardia and auricular fibrillation. The contribution of Lewis' research to bridge Einthoven’s research was vital and can be understandable by Einthoven’s statements after he was awarded the Nobel Prize in 1924. 

In his Nobel lecture, Einthoven stated about Lewis, ``\textit{I owe you so much. Without your steady and excellent work to which you have devoted a great part of your life there would have been in all probability no question of a Nobel prize for me. You have given to Medicine at least as much as I have}'' \cite{Nobellecture}. Lewis continued working as the world’s leading electrocardiographer while  Einthoven investigated the theoretical bases for  Electrocardiography. 

\begingroup
\centering
\begin{figure}[htbp]
	\centering
	\subfigure[Atrial Fibrillation waveforms by Thomas Lewis \cite{picture}  ]{\includegraphics[width=0.4\textwidth]{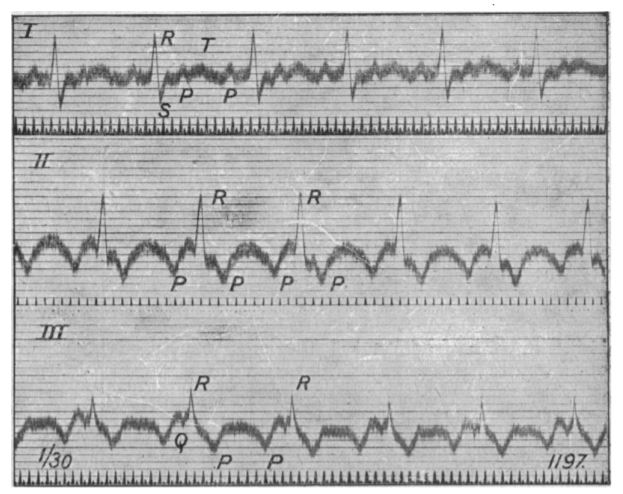}}\label{tl1}
	\subfigure[ECG Waves of Mother and Feotal\cite{picture}] {\includegraphics[width=0.45\textwidth]{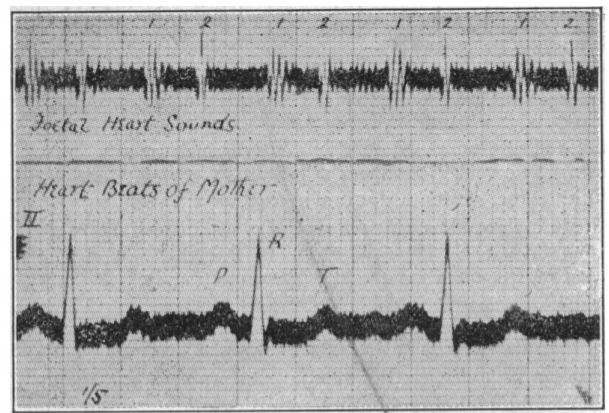}}\label{tl2}
	\caption{ ECG Signals studied by Thomas Lewis (Picture Credit:\cite{picture})} 
	\label{tl}
\end{figure}

\endgroup

The clinical features of Myocardial Infarction were first published in 1910 by Russian Physicians W P. Obrastzow and N. D. Straschesko. They reported two main findings: prolonged chest discomfort and persistent dyspnea\footnote{Shortness of breath or difficulty in breathing} however, these features were not based on ECG tracings. Further, the electrocardiographic features for various diseases were observed as listed below.
\begin{itemize}
	\item 1917-  Electrocardiographic features for acute myocardial infarction were first published by Oppenheimer and Marcus Rothschild \cite{myocardial}.
	
	\item  1920- Harold Pardee reported ST-segment elevation in Lead II, Lead III and T wave inversion features of Myocardial Infarction \cite{myocardial2}.
	
	\item 1924-Woldemar Morbitz found out two different types of second-order Atrioventricular (AV)  blocks that are named after him, known as Morbitz type I and type 2 blocks \cite{Morbitz}.
	
	\item 1930- WPW syndrome was described by scientists Wolf, Parkinson and White named after the scientists in which Bundle branch block with short PR interval was discussed \cite{WPW}.
	
	\item 1931-1932- Charles Wolferth and Francis Wood reported the electrocardiographic features for the Angina Pectoris\footnote{Angina pectoris is the medical term for chest pain or discomfort due to coronary heart disease. It occurs due to narrowing of arteries due to blockage and also known as ischemia} after moderate exercise \cite{wood1, wood2}.
	
	\item 1935- Sylvester McGinn and Paul D. White find features for the cardiac condition, acute pulmonary embolism \footnote{It is a blockage of an artery in the lungs.} \cite{Ginn}.
	
	\item 1939- Richard Langendorf obtained the ECG features for Atrial Infarction \cite{lang}.
	
	\item 1942-  Arthur Master, Friedman Rudolph and Dack Simon standardized the two step exercise test also known as the Master two-step for cardiac function \cite{Arthur}.
	
	\item 1944- Young and Koenig reported the PR Segment deviations for Atrial Infarction condition \cite{pr}
\end{itemize}

The development of such Electrocardiographic features is till date continued.

\subsection{Evolution of ECG Leads}
Developmental activities were also going to find out the best materials for electrodes. Electrodes are connected to various specific locations on the human body. Electrodes are conductive pads that enable the recording of the electrical activity of human heart.
An 'ECG lead' can be obtained by analyzing the various electrode signals and obtained by considering different positions of the electrodes. It provides different viewpoints to measure the heart's electrical activity and is similar to clicking a picture of heart from different angles to get a better understanding by the physicians. 

Leads may be unipolar or bipolar in nature. In the unipolar leads, the potential difference between any specific electrode and ground terminal is considered.
In Bipolar leads, the difference between two electrodes' signal is considered with reference to the ground terminal.  Unipolar leads provide the horizontal view of the heart, and bipolar leads provide the heart's frontal view. 

\begin{figure}[htb!]
	\centering
	\includegraphics[scale=0.18]{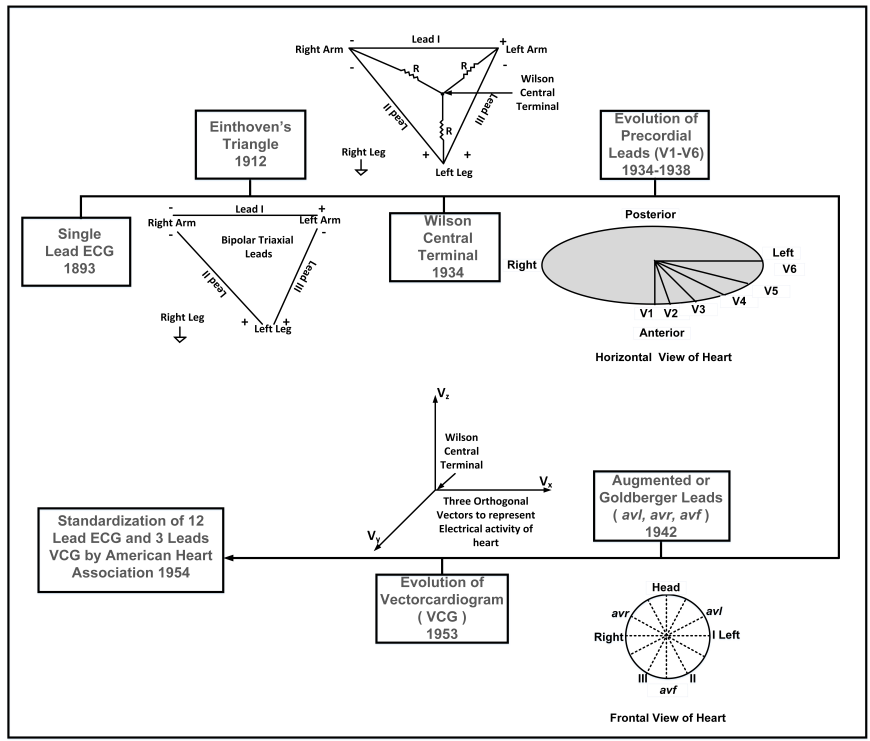}
	\caption{ Development of ECG Leads and VCG}
	\label{leads}
\end{figure}

During 1893, Einthoven first used the term EKG and studied the graphs using the capillary electrometer. Later he built a string galvanometer based on a 3-electrode EKG machine in 1902. In the year 1912, Einthoven mathematically reported the Einthoven's triangle  \cite{Einth4}. This became the basis for future EKG, Vectorcardiography (VCG) and development of Electrodes and Leads for ECG acquisition (see Fig.\ref{leads}) that is being used till date.  Table \ref{LE} shows the corresponding number of electrodes and leads for ECG and VCG schemes.

\begin{table}
	\centering
	\caption{ ECG and VCG Leads and Electrodes}
	\label{LE}
	\fontsize{6}{8}\selectfont
	\begin{tabular}{c c c c c c }
		\rowcolor{lightgray}\  \textbf{Signal} & \textbf{Leads} &\textbf{(Bipolar/ } & \textbf{ Number of } & \textbf{Location of }   \\
		\rowcolor{lightgray}\  \textbf{Acquisition} &  &\textbf{Unipolar/} & \textbf{ Electrodes} & \textbf{ Electrodes}   \\
		\rowcolor{lightgray}\  &  &\textbf{Vectors)} &  &  \\
		\hline
		\multirow{ 5}{*}{Standard 12 Lead ECG} & I  & Bipolar & 3 & RL, LA, LL  \\
		& I, II, III &  Bipolar & 4 & RA, LA, LL, RL  \\ 
		& \textit{avl}, \textit{avr}, \textit{avf} & Unipolar & 4 & RA, LA, LL,RL  \\ 
		& (V1, V2, V3 & Unipolar & 6 & Different anatomical  \\ 
		& V4, V5, V6) & & & sites on chest \\
		
		\hline
		\ \multirow{ 3}{*} {Vectorcardiogram} &$V_{X}, V_{Y}, V_{Z}$ & Vectors & 7 & 5 at the transverse \\      
		\ &&&&  plane of chest,\\
		\ &&&& BN and LL \\ 
		\hline
	\end{tabular}
	\footnotesize { $^{LL}$ Left Leg, $^{LA}$ Left Arm, $^{RA}$ Right Arm,  $^{RL}$ Right Leg,  $^{BN}$  Back of Neck }\\
\end{table}

\begin{figure}[htb!]
	\centering
	\includegraphics[scale=0.25]{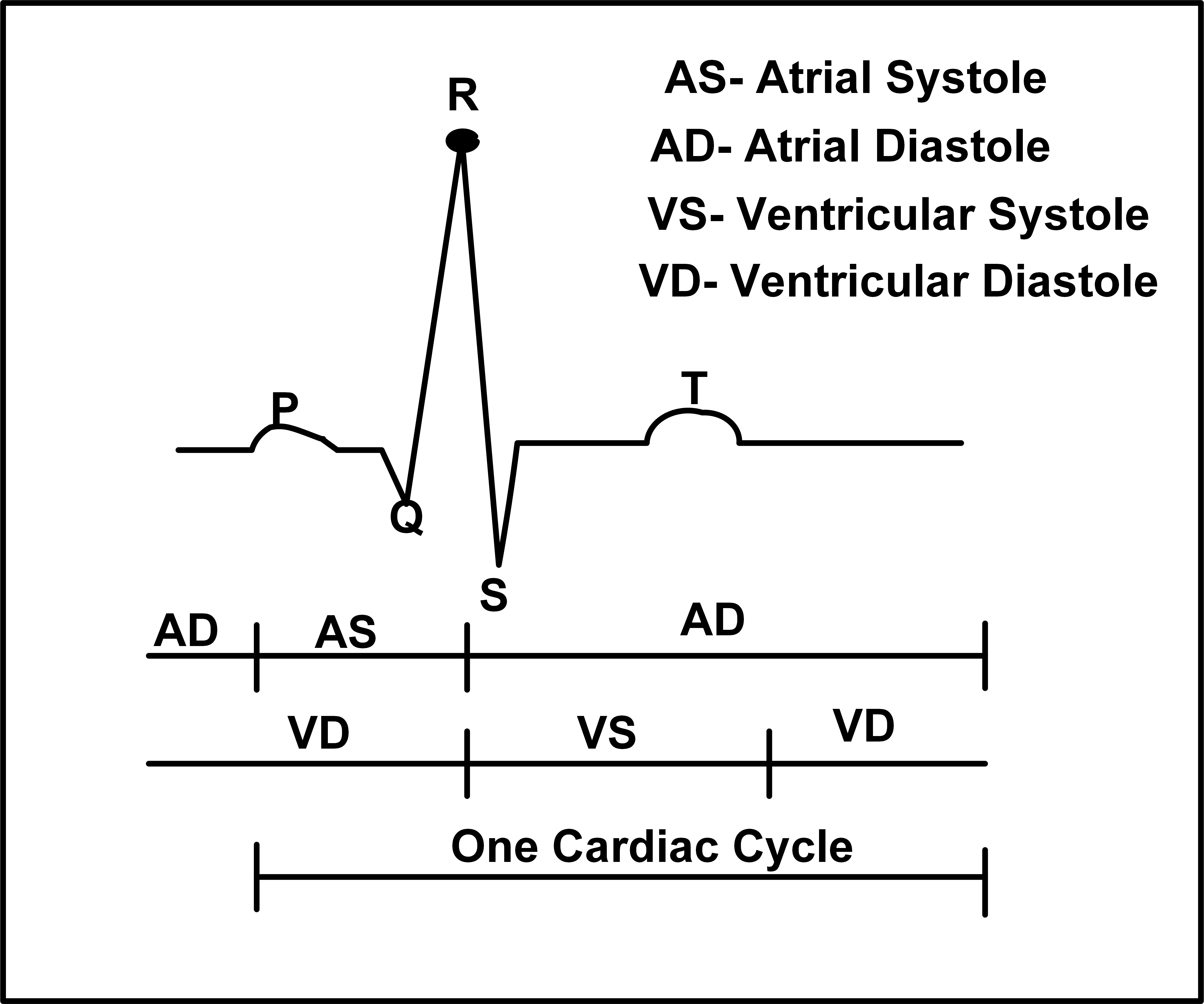}
	\caption{Representation of a Cardiac Cycle}
	\label{cc}
\end{figure}

In 1934, Frank Wilson defined an `indifferent electrode' that was later known as `Wilson Central Terminal' by connecting the right arm, left leg and left arm with resistances typically $5K\Omega$ \cite{WCT}. Wilson Central Terminal is an artificially constructed reference for electrocardiography, which is assumed to be at zero potential and steady during the cardiac cycle so that the reference point for unipolar potential remains fixed. It worked as a ground terminal for other unipolar leads. The events that occur during each heartbeat are termed a cardiac cycle that can be divided into two parts: a period of relaxation known as diastole and a period of contraction known as systole. A cardiac cycle on an ECG signal is shown in Fig. \ref{cc}. 

In 1938, the American Heart Association and the Cardiac Society of Great Britain published their recommendation for recording the exploring lead from six sites named V1 through V6 across the precordium \cite{WCT}.

 Later, Emanuel Goldberger extended the Wilson Central Terminal with Augmented Unipolar Leads (\textit{avl},\textit{avr} and \textit{avf}) also known as Goldberger leads for obtaining a detailed view of the frontal plane \cite{Goldberger}. Further, in 1953 the general theory of heart vector projection was presented by Earnst Frank \cite{Frank} that provided a mathematical framework where three vectors determined the person's complete cardiac health. This confirmed the robustness of the methods used then with a mathematical validation. In the following year in 1954, the American Heart Association published their recommendation for standardization of 12-lead Electrocardiogram and Vectorcardiogram \cite{AHA}. Till date, 12 Lead ECG and 3- Lead VCG continues to be the standard of ECG measurement systems.
 
 \subsubsection{Types of ECG Electrodes}
  ECG electrodes are the conductors to obtain the electrical signals of heart activity from the body and are connected to specific body locations.  During the invention of the String Galvanometer, Einthoven utilized the saline water buckets as Electrodes. Nowadays,   Silver - Silver Chloride (Ag- AgCl) are mostly used to obtain ECG signals. These electrodes are connected with the electrolyte gel on the skin to increase the conductivity, hence also known as wet electrodes. An electrolyte is usually composed of a salt solution gel material. Ag- AgCl electrodes are widely used in conventional schemes as these type of electrodes provide a high signal to noise ratio in the obtained signals, but there are some disadvantages associated with these types of electrodes.
  Some patients seem to be allergic to these gels and while the hairs on the skin make it difficult to apply for some cases. Also, the wet electrodes are not comfortable for long term ECG monitoring \cite{elreview}. 
  
  Electrodes can be categorized as active and passive electrodes. In the active electrodes, the pre-amplification module is immediately after the conductive material between the skin and the electrode and is present to enhance the signal to noise ratio of the signal. The passive electrodes just provide the direct connection between the metal layer and the processing unit. 
  Various other types of electrodes are also reported in the literature and are compared with the conventional wet Ag-AgCl electrodes \cite{el1, el2, el3, el4, el5, el6, el7, el8,el9, el10,el11,el12}, . A list of other types of electrodes recently reported in the literature is given- below:

  Gel less ECG Electrode \cite{el1}: This work discussed the grip-style dry electrodes for ECG measurement during physical activity and present an innovative design for a portable ECG amplifier that mitigates some of the pre-identified issues of electrodes.
  
  Capacitive Electrodes for noncontact ECG monitoring \cite{el2, el9}:The noncontact capacitive electrodes can obtain ECG signals through clothes and implemented with the real-time denoising algorithm.  
  
  Carbon Nanotube (CNT)/Polydimethylsiloxane (PDMS) composite-based dry electrodes \cite{el3}: In this work, the CNT/ PDMS composite-based dry ECG
  electrodes were readily connected to the conventional ECG
  devices, and showed its long-term wearable monitoring capability
  and robustness to motion and sweat.

  Esophageal Electrodes for long term monitoring based on Titanium Nitride (TiN) and iridium oxide (IrOx) \cite{el4}: This work discussed the advantages of esophageal electrodes over wet and dry electrodes for long term monitoring without the need of electrolyte gels.
  The TiN and IrOx are identified as suitable materials for esophageal electrodes that are superior to the standardized surface skin-electrode concerning signal distortions, and thus, it might help prolong conventional ECG recordings maintaining high-quality signals. 
  
  Underwater Electrodes based on Carbon Black Powder (CB) and Polydimethylsiloxane (PDMS) \cite{el5}: In this work, hydrophobic electrodes that provide
  all morphological waveforms without distortion of an ECG
  signal for both dry and water-immersed conditions was discussed.
  
  Dry Metal Electrodes \cite{el6} :The wearable multi-lead
  electrocardiogram (ECG) recorder suggested that dry metal electrodes provide a
  comfortable sensation, less skin-irritating, easy clean surfaces,
  reusable capability, and more durability compared to conventional ECG electrodes.
  
  Dry textile-based electrodes, needle array electrode and a
  silver coated surface electrodes \cite{el7}: The three types of dry electrodes viz. textile electrodes, needle array electrode and a Silver coated surface electrodes were fabricated and tested for acquisition of ECG signals. The dry electrodes were fabricated as
  active electrodes and recording from the dry electrodes are compared to that of the wet	electrodes.
  
  Carbon Based Electrodes for Wearable Applications \cite{el8}:
  The flexible dry electrodes for long-term biosignal monitoring was designed by mixing carbon nanofibers (CNFs) in biocompatible-elastomer (MED6015).
  
  Silver Nanowire based Dry Electrode \cite{el10}:   	
  The silver nanowire (AgNW)-based dry electrodes were fabricated for noninvasive and wearable ECG sensing.
  
  Garment type electrode \cite{el11}: The multi-channel telemeter and
  garment-type electrodes were developed that exhibited a sufficient R-wave
  detection rate in four positions
  
  Poly(3,4-ethylenedioxythiophene) Polystyrene Sulfonate (PEDOT:PSS) and PDMS coated cotton fabric electrode \cite{el12}: A flexible electroconductive textile material was developed by coating PEDOT:PSS/PDMS on cotton fabric via flat screen printing. The coated fabric was utilized as ECG electrodes and compared with the conventional electrodes. 
      
\section{  Cardiac Health Monitoring Schemes}
Other streams of cardiac health monitoring that are also of interest to researchers are PhonoCardioGraphy (PCG), BallistoCardioGraphy (BCG), ApexCardioGraphy (ACG), SeismoCardioGraphy (SCG) and KinetoCardioGraphy (KCG).

Phonocardiogram is a plot of high fidelity recording of the sounds made by the heart. Monitoring and recording equipment for PCG was developed during  1930- 1940s and was standardized around 1950 \cite{PCG1}. PCG originated in an attempt to time the occurrence of heart sounds in a cardiac cycle. The acquisition system of PCG, similar to ECG, consists of low noise amplifiers and filters. As this review focuses only on the ECG and related signal processing schemes, the interested reader may refer to more details regarding PCG in \cite{PCG1, PCG2, PCG3}. Ballistocardiogram measures the Ballistic forces (Mechanical Forces) generated by the heart and is a non-invasive method based on the measurement of the body motion generated by the blood's ejection at each cardiac cycle \cite{BCG}.  Apexcardiogram was first described by Marey in 1878 \cite{ACG1}. The curves of the apexcardiogram display all consecutive phases of the cardiac cycle; contraction-and-emptying and relaxation-and-filling. The apex cardiogram's waveform is caused primarily by movements of the left ventricle against the chest wall. Thus, it is a translation of the sequence of hemodynamic events occurring in the underlying left ventricle \cite{ACG2}. Seismocardiogram (SCG) is the recording of body vibrations induced by the heartbeat. SCG contains information on cardiac mechanics, in particular, heart sounds and cardiac outputs \cite{SCG}. Kinetocardiography (KCG) records indicate movements as the result of the motions of the heart and utilize only the low-frequency motions (0--30 Hz)\cite{KCG}.

Though these methods by themselves are interesting, none evolved to match the process of ECG methods.

\section{Computerized Automated Detections on ECG for Cardiac Health Monitoring in Early Years [60's -90's]}

By 1961, Norman Jeff Holter had designed the Holter Monitor  \cite{Holters} for continuous ECG monitoring in hospitals.  Holter designed a backpack recorder that weighed 75 lbs and it was able to record and transmit ECG signals to the hospitals for further evaluation. It became a landmark invention for the automatic detection and transmission of biomedical signals, specifically ECG signals. 

A new era had ushered for automated computerized detection. The first automated classification on 20 clinically normal individuals on magnetic tape recorders to store a 1-minute recording of each subject \cite{Caceres} were utilized. In 1965, average transient computing based on average response computing techniques were presented to extract the ECG signal from the noisy records such as exercise stress test results in \cite{Raut}. In 1966, the system was developed for online computer monitoring of critically ill patients \cite{Jensen}. This system was able to provide continuous monitoring for various health parameters such as ECG, systolic and diastolic blood pressure, pulse rate, temperature readings at various parts of the body, manual inputs from the user etc.  within the hospital setting due to the high form factor of the system. The system was used at the California Shock Research Unit for clinical management of serioulsly ill patients. The system provided online acquisition, processing and display of the data.  In 1967, Vector Cardio Graph (VCG) was used to separate normal and Left Ventricular Hypertrophy (LVH) on subjects \cite{Klingeman}. Four different techniques were utilized for categorization, such as sum of amplitude measurement, vector differences, weighted vector differences and class separating differences. Two hundred subjects’ samples were utilized for obtaining the results out of which 100 were LVH Samples and 100 were normal subjects (case-control study). 
In the similar year another work \cite{Brody} was presented to analyze the normal subjects' frontal leads to expand the conventional amplitude and time base factor. 

By 1968 advanced mathematical concepts from signals and systems viz Contour Analysis were used to categorize the normal and abnormal ECG Rhythms \cite{Pordy}. Around 2000 samples were subjected to an automated contour plot and the results were compared with the physician’s results to determine the accuracy of the technique.

By 1970, the online Real Time algorithms for ECG waveforms were started to developed \cite{Haywood}, that provided the information whenever the  values were exceeded beyond the predefined limits for a single lead ECG signal.

In 1971, both 12 lead ECG and 3 Lead VCG techniques were considered standard methods of monitoring and around 1100 patients for the automated detections were obtained \cite{ Macfarlane, Macfarlane2} were taken. This study suggested no appreciable differences from automated techniques obtained from both 3 Lead VCG and 12 Lead ECG data inputs. It also concluded that 3 Lead VCG data was sufficient for automated detections as it saved the computing time and emphasized VCG as preferable for automated detections. Although VCG became famous for automated detections, it never became popular for physicians due to the nonstandardized lead configuration, and the other reason is medical doctors are accustomed to using 12 lead ECG in clinical applications \cite{vcg}. VCG can be obtained by 12 lead ECG signals but this field also requires more attention from the researchers.
  
In 1972, the computer analysis of rest and exercise electrocardiogram based on three inputs frank leads or six bipolar leads were presented in \cite{Agarwal}. The authors utilized the Dalhousie ECG analysis program with the statistical properties of ECG signal. In 1973,  a research based on clustering techniques \cite{Swenne} for pattern recognition. The categorization was based on the clusters defined by the human operators. If the values falls outside the boundary limits of clusters the human invention was requested. The system was semiautomated and classified the QRS complexes of the ECG signal. In 1974, another research work \cite{Yanowitz} was published for  QRS and premature ventricular beat detection for a continuous real time ECG signal. However, the authors were not convinced for its widespread use and specified that it was limited for some specific research purposes. 
  
Around the similar year, $\mu$processors were made available for automatic analysis in biomedical engineering. The first reported use of $\mu$processor for Ambulatory ECG monitoring is believed to be around 1976 \cite{Walters, Schluter}.  These were $\mu$processors powered bedside ECG monitors and became an important milestone in automated ECG methods due to transformation in the research methodology. Various papers were reported \cite{Webster, Tompkins, Thakor, Abenstein, Mueller, Agarwal} during the same period that provided the $\mu$processors based systems for Ambulatory ECG Monitoring.

Until 1980's, all these automated detection algorithms and methods were based on separate databases. So, there was a need for a standard database as the results were invariably data-dependent. Literature \cite{Hermes, Mark, Thakor2} suggests that various attempts to obtain the Gold Standard database for comparison begun. Around 1983, MIT and Beth Israel Hospital Arrhythmia Laboratory released the data obtained from Holter tapes patients between 1975 and 1979 \cite{Mark}. This later became a standard database and is used till date.  The databases \cite{db1, db2, db3, db4, db5, db6, db7, db9, db10, db11, db12, db13, db15, db17, db18, db19, db20, db21} available on Physiobank for automated processing schemes comparisons are shown in Table \ref{db}.  PhysioBank is a large and growing archive of well-characterized digital recordings of physiologic signals and related data for use by the biomedical research community. In this paper, only ECG and relevant signals databases are considered. The databases provide the annotations for quantification of ECG waves (Truth values for P, QRS and T waves to be compared with automated generated results) and the diagnosis of various diseases confirmed by cardiologists. Except this, some of the databases also provide information for compression tests and signal to noise ratio information of the signal. These standard databases continue to be the gold standard of ECG processing research.

\begin{table}
	\centering
	\caption{Various ECG Databases available on Physionet}
	
	\label{db}
	\fontsize{6}{8}\selectfont
	\begin{tabular}{c c c c c}
		\rowcolor{lightgray}\ \textbf{Ref.} &\textbf{Database} & \textbf{Specifications} & \textbf{Number of}  & \textbf{Abbreviations} \\
		\rowcolor{lightgray} &&& \textbf{Records} & \\
		\hline
		\ \cite{db1} & MIT-BIH   & 2- Channel Ambulatory  & 48 & MITDB\\
		\ & Arrhythmia & ECG with R peak & \\
		\ & & Annotations \\
		\hline
		\ \cite{db2} & MIT-BIH Arrhythmia  & Contains P-wave& 12 & PWAVE\\
		\ & P-Wave Annotations &  annotations for 12 & \\
		\ &&  signals from MITDB\\
		\hline
		\ \cite{db3} & MIT-BIH Atrial & 25 long-term ECG   & 25 & AFDB \\
		\ &  Fibrillation & recordings with & \\
		\ & & Atrial Fibrillation \\
		\hline
		\ \cite{db4} & MIT-BIH ECG   &  Short ECG recordings & 168 & CDB \\
		\ & Compression Test  &  to pose challenges  & \\
		\ & & for ECG compressors & \\
		
		\hline
		\ \cite{db5} & MIT-BIH Long-  & Long-term ECG   & 7 & LTDB\\
		\ & Term ECG & recordings with   & \\
		\ & & manually reviewed \\
		\ & & beat annotations \\
		
		\hline
		\ \cite{db6} & MIT-BIH Malignant  & Recordings of Ventricular & 22  & VFDB\\
		\ & Ventricular, Ectopy &  Fibrillation, Flutter& \\
		\ && and Tachycardia  \\
		
		\hline
		\ \cite{db7} & MIT-BIH Noise  & Recordings of & 12+3 & NSTDB\\
		\ & Stress Test &  different SNR data  & \\
		\ &&and Typical noises in\\
		\ & & Ambulatory ECG & \\
		
		\hline
		\ \cite{db5} & MIT-BIH Normal   & Long term Normal  & 18 & NSRDB\\ 
		\ & Sinus Rhythm & Sinus Rhythm \\
		
		\hline
		\ \cite{db9} & MIT-BIH  & Multiple physiologic  & 18 & SLPDB\\
		\ &  Polysomnographic & signals during sleep \\ 
		
		\hline
		\ \cite{db10} & MIT-BIH  &  Recordings of varying & 28 & STDB\\
		\ & ST Change & length during exercise   & & \\
		\ & & stress tests and with\\ 
		\ & & transient ST depressions & & \\
		
		\hline
		\ \cite{db11} & MIT-BIH  &  Recordings of & 78 & SVDB\\ 
		\ & Supraventricular  & Supraventricular & & \\
		\ & Arrhythmia &  arrhythmias\\
		\hline
		
		\ \cite{db12} & QT  & With onset, peak,  & 100 & QTDB\\
		\ & & and end markers  \\
		\  & &  for P, QRS, T, and  U \\
		\ && waves annotations \\
		
		\hline
		
		\ \cite{db13} & Physikalisch-Technische  & 15 Leads data for    & 549  & PTB\\
		\ & Bundesanstalt & various cardiac conditions   & &\\
		\ && specifically MI\\
		\hline
		\ \cite{db5} & St Petersburg  & 12 standard leads  & 75 & INCARTDB \\
		\ &  INCART 12-lead   & annotated recordings   & \\
		\ & Arrhythmia & extracted from 32  \\
		\ & & Holter records \\ 
		\hline
		
		\ \cite{db15} & European & ST segment change  & 90 & EDB\\
		\ &  ST-T  & and T wave change  & & \\
		\  & & episodes included \\
		
		\hline
		\ \cite{db5} & Common Standard   & collection of short & 1000 & CSE\\
		\  & of  &  (12- or 15-lead)  recordings & \\
		\ &  Electrocardiography\\
		\hline
		\ \cite{db17} & Combined measurement  & ECG, SCG and  & 20 & CEBSDB \\ 
		\ & of ECG, Breathing  &  Breathing signals  \\
		\ & \&] Seismocardiograms & of healthy subjects\\
		\hline
		
		\ \cite{db18} & WECG  & Wrist ECG of  & 30 & WECG\\ 
		\ & & healthy subjects \\
		
		\hline
		\ \cite{db19} & FANTASIA  &  ECG and respiration  & 40 & FANTASIA\\
		\ & &  signals during  \\
		\ & & supine resting \\
		\hline
		\ \cite {db20} & Apnea-ECG  & Annotated nighttime  & 90 & APNEA-ECG\\
		\ &  &  ECG recordings\\
		\hline
		
		\ \cite{db21} & PAF Prediction & consists of training  & 50-Training & AFPDB\\
		\ & Challenge  & and testing  data for  & 50- Testing \\
		\ & & Proxymal Atrial \\
		\ & & Fibrillation detection\\
		\hline
	\end{tabular}
\end{table}

Research related to Holter Tapes' automated analysis with $\mu$computers and Automated Holter Scanning were published during 1983 \cite{Ahlstrom}. The system consisted of two  $\mu$computers to detect QRS durations for arrhythmias of 24 hours recorded on Holter tapes. It determined the heart rate variability and PVC counts, a method used till date. Around the same year another method of QRS complex detection was presented in \cite{Nygard}. In this method the QRS complex was represented by single positive pulse alongwith onset and end of it, by a dynamic threshold technique by utilizing the time domain features. But, the results of the method was provided on any simulated ECG data as software based technique. 

A portable  $\mu$computer based Arrhythmia Monitor was designed \cite{Thakor3} for storing 16 \textit{seconds} arrhythmia intervals. The major difference of this system with Holter tapes was that it did not store any normal rhythm data and was advantageous in terms of memory utilization.  The system was able to provide continuous and long term monitoring for high-risk patients.  In 1987, filter design was illustrated in \cite{Thakorfilter} for biomedical signal processing techniques. The filters were implemented on the ECG signals and quantisation of filter coefficients were used to design various different filters with the its implementaion on 8 bit $\mu$processor.

These $\mu$processor-based systems became the interest of researchers with databases available to them and the evolution of data compression techniques. The processing of various data compression techniques also evolved rapidly as processing the 12 Lead ECG data for automated methods were computationally expensive. The compression techniques further minimized redundant information present in the original signal and helped the system practically feasible. These methods are believed to have early beginnings during 1968 and continued in later years (1984- 1992) as cited in various works \cite{Tompkinsbook, Abenstein2, Bohs, Cox, Hamilton}.

Various data compression algorithms as Turning Point Algorithms \cite{tp}, AZTEC Algorithm \cite{aztec}, CORTES algorithm\cite{cortes}, Fast walsh Transform \cite{fastwalsh}, FAN algorithm \cite{fan} and SLOPE algorithm \cite{slope} are discussed. One of the main concerns in biomedical data reconstruction was the clinical acceptability of these signals.
During the data compression, the requirement for lossless information and removal of repeated or redundant signals proved a challenge. The data reduction techniques provided viable options for storing or processing large amounts of data with lower storage requirements and became successful. 

Following this, Pan and Tompkins proposed the seminal algorithm for QRS detection \cite{Tompkins2} for normal and abnormal waveforms. This algorithm provided accuracy of more than 99\% for QRS detection and revolutionized the means for arrhythmia monitoring. The algorithm also provided the ideal means for heart rate variability measurements that provided real-time processing and reporting various cardiac conditions and diseases. In 1987, another research \cite{HRVAR} provided the preliminary heart rate variability (HRV) analysis by using the autoregressive modelling techniques and power spectral density estimates. For the QRS detection, it followed the classical technique by obtaining the derivative of the ECG signal followed by a adaptive thresholding. After obtaining the R-R interval information, it used to discrimate the normal and pathological subjects  by utilizing the autoregressive modelling and power spectral density estimates. In 1988, two methods for detecting the QRS complexes were discussed in \cite{transformation} based on the length transformation and energy transformation of the signal. In both the methods QRS complexes of the signals were enhanced and other components of the signal were supressed significantly and detection accuracy for QRS complexes were found out be over 99\%. 

Around the year 1989, research on connectionist systems, better known as neural networks for diagnostic purposes, was proposed. Neural networks were first used for ECG signal processing during 1990 for diagnostics \cite{ Neural1, Neural2, Neural3, Neural4, Neural5, Neural6, Neural7}, categorization and QRS detections and proved interesting. The application of neural networks also proved to be advantageous in classifications and detections with extended computations. Over the years, such artificial intelligence algorithms were extended towards categorizing normal and abnormal waveforms and pattern matching. During 1992, detection of QRS complexes for very noisy signals were demonstrated using neural networks \cite{Neural5}. In this work, a multilayer perceptron neural network was used as an adaptive whitening filter instead of a normal linear filter. 
Another work on the QRS template matching was updated by ANN recognition algorithm was discussed in \cite{Neural6}.  Several hidden layers in multilayer perceptron with eigenvalue decomposition method to classify the signals available in MIT/BIH database was provided in \cite{Neural7} and the technique was also patient adaptable. Although the neural networks provided better detection accuracy over the conventional classification (based on thresholding and empirical values) but the computations requirements for such systems were high and often difficult to realize on customized hardware.

\begin{figure*}[htb!]
	\centering
	\includegraphics[scale=0.30]{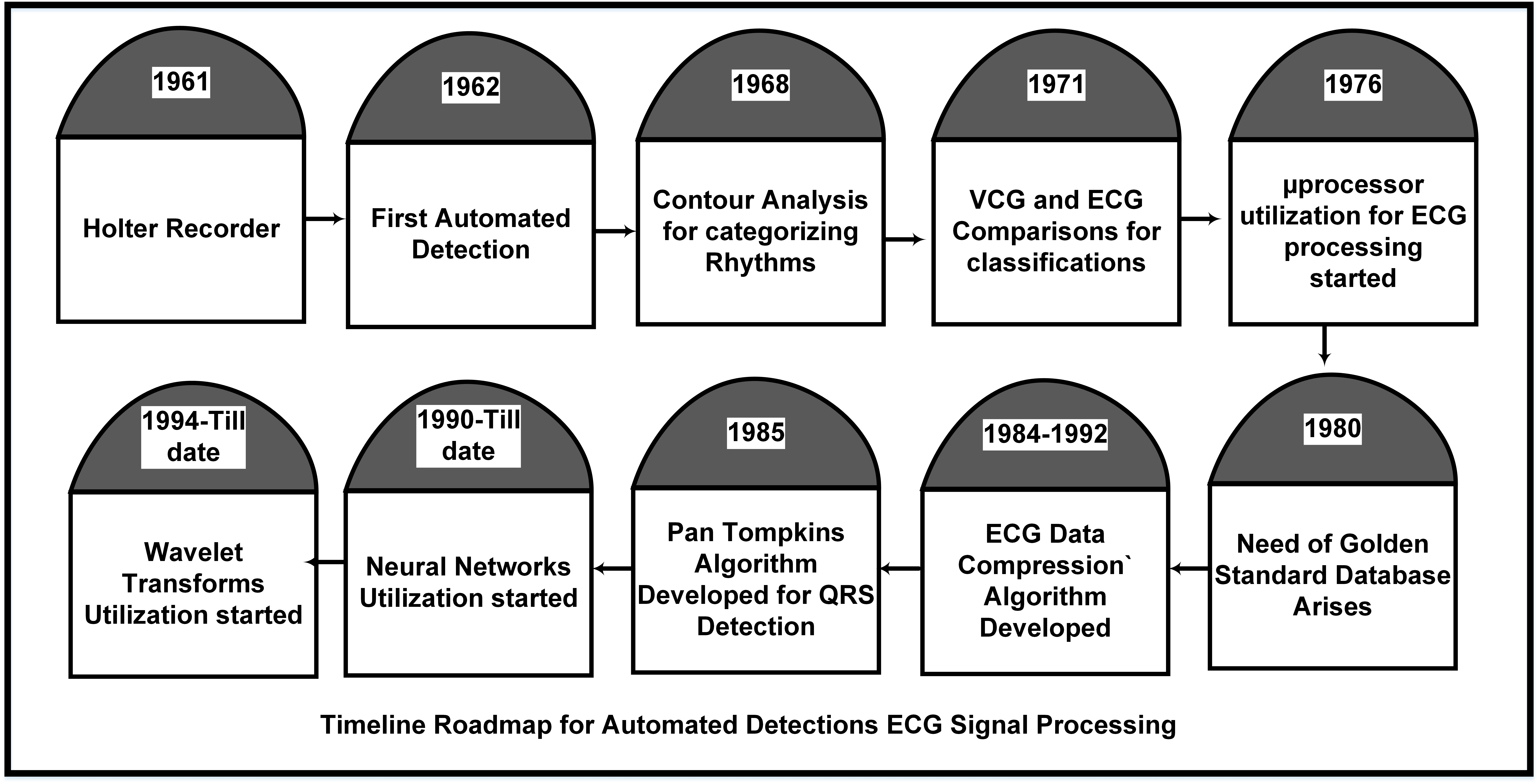}
	\caption{ Milestones in ECG Signal Processing till the Year 2000 }
	\label{f4}
\end{figure*}

In \cite{Wavelet0}, authors discussed the utilization of Wavelet Transform (WT) for ECG analysis and it's compression techniques. The research presented preliminary investigation into its application to the study of both ECG and heart rate variability data.
Futher, WTs were also studied independently that provided time and frequency analysis for the ECG signals discussed in \cite{Wavelet1, Wavelet2, Wavelet3, Wavelet4, Wavelet5}. The authors suggested that wavelet transforms' efficiency measures were comparatively higher than conventional methods \cite{Wavelet1, Wavelet2}.  
The timeline for the signal processing is shown in Fig. \ref{f4}.

\section{ECG Signal Processing in Recent Years[90-Till date]}
\begin{figure*}[htb!]
	\centering
	\includegraphics[scale=0.20]{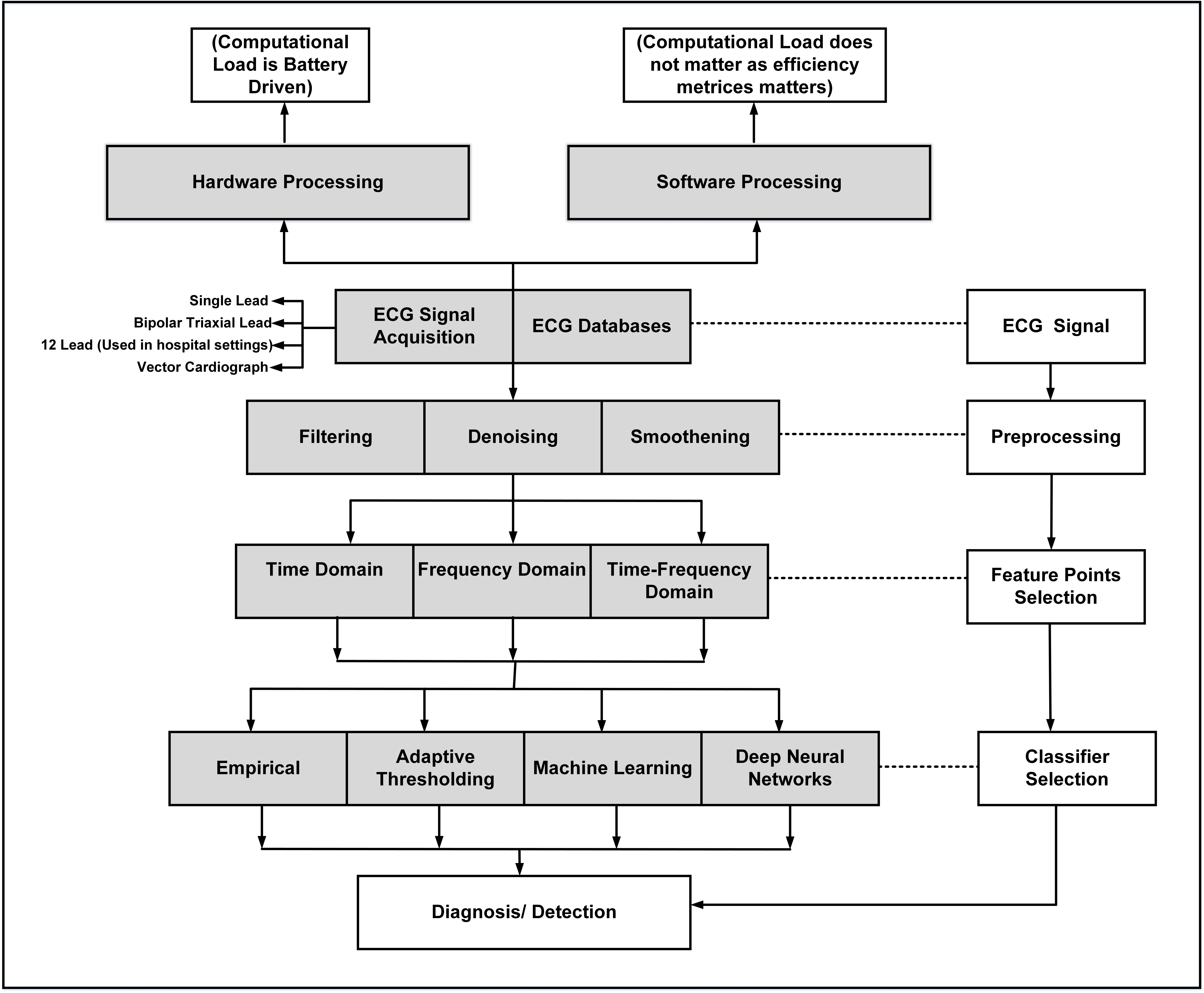}
	\caption{ ECG Signal Processing}
	\label{f5}
\end{figure*}

Fig. \ref{f5} shows various steps of standard ECG signal processing consists of preprocessing, feature points selection and classifiers. The preprocessing step is the initial step in signal processing before feature selection. The preprocessing stages consist of denoising the ECG signal affected by various kinds of noises such as baseline wandering, power line interference, electromyogram noise \textit{etc.}\cite{Noise}. In some of these cases, filters are employed at this level to consider a particular band of frequencies. Introduction of the preprocessing stage before feature selection leads to more accurate results. The features of the signal may have temporal, spectral, time-frequency information. In some of the cases, the statistical information regarding the signal is also considered. 

After the feature selection, classifiers are implemented to categorize the signals. These classifiers may be empirical, thresholding, machine learning or a deep neural network. Based on the results of classifiers, detection of the disease or diagnosis is usually done.
In literature , time domain \cite{td1, td2, td3, td11, td4, td5, td6, td7, td8, td9, td10}, frequency domain methods \cite {fd1, fd2, fd4, fd5, fd7, fd8, fd9, fd10, fd11, fd12, fd13} or time frequency domain methods \cite{tfd1, tfd2, tfd3, tfd4, tfd5, tfd6, tfd7, tfd8, tfd9, tfd10, tfd11, tfd12, tfd13, tfd14}  are classified with the machine learning or deep neural network approaches. In this report, the papers are classified according to feature point selection or the classifiers scheme discussed in the work. 

The signal processing can be done after using the hardware and software only systems. In the software processing schemes, the computational load does not pose a constraint on the system. Therefore, complex processing schemes are often used to obtain results that provide comparatively more accurate results. However, in the Real-Time system for resource-constrained regions, the computational load does matter and poses a real challenge for the battery-driven systems, deployment and relevance. Hence, for automated detections, there must always be a tradeoff between performance and battery requirements. The performance matrices of the system is defined in terms of True Positives (TP), False Positives (FP), True Negatives (TN), False Negatives (FN), Accuracy(Acc), False Detection Rate (FDR), Sensitivity (Se), Specificity (Sp), Positive Predictivity (PPV), Error Rate (ER), Efficiency of Recognition (EOR), Classification Rate (CR), Classification Error Rate (CER), False Acceptance Rate (FAR), False Rejection Rate (FRR) and Detection Error Rate (DER) etc \cite{td1, td2, td3, td11, td4, td5, td6, td7, td8, td9, td10, fd1, fd2, fd4, fd7, fd8, fd9, fd10, fd11, fd12, fd13, tfd1, tfd2, tfd3, tfd4, tfd5, tfd6, tfd7, tfd8, tfd9, tfd10, tfd11, tfd12, tfd13, tfd14 }.

\subsection{Feature Points Selection}
Feature point selection is a crucial stage as it requires the tradeoff between the system complexity, accuracy and battery requirements for the system. Processing techniques may vary from the simple morphological features capturing to complex transformations. Features may be from time domain, frequency domain, or a mix of both as in time-frequency domain. Researches related to the field are listed and compared in the following subsections. 

\subsubsection{Time Domain}
The time domain processing or the temporal domain feature extraction allows the processing in discrete samples. With the evolution of $\mu$processors around the year 1974, the conversion of continuous signals to discrete ones became usual, so the signals also were usable by $\mu$processors. It was easier to convert the discrete signals again back to analog \cite{Tompkinsbook}. Authors in \cite{td1-1}
discussed the Noise filtering, QRS detection, wave delineation and data compression of the ECG signal in time domain. As shown in Table \ref{Timedomain}, in \cite{Tompkins2}  the Real-Time QRS detection algorithm based upon digital analyses of slope, amplitude, and width is discussed and was validated using MITDB. The main disadvantage of this was that it required back processing of ECG signals and a significant memory for real-time processing. In \cite{td1}, authors discussed two algorithms for QRS detection; the first algorithm detects the current beat while the second algorithm has an RR interval analysis component. 

 Authors utilized the difference operation method for QRS complex detection in two stages in \cite{td2}. The first stage was to find the R point by applying the difference equation to an ECG signal and the second stage looked at Q and S points based on the R point to find the QRS complex. By doing so, Q and S point detection efficiency was dependent on the R peak detection algorithm. 
In \cite{td3}, authors utilized the squared double-difference signal of ECG signals to detect the R peaks and the results are provided only for the normal cases. In \cite{td11}, authors utilized a time-domain morphology and gradient-based approach based on a combination of extrema detection and slope information, using adaptive thresholding for ECG features extraction. The main limitation of this algorithm was its robustness not tested against baseline wandering variations. In \cite{td4, td5}, authors determined the P and T waves of ECG signal using the two moving average filters that provided the dynamic event-related thresholding by utilizing the signal's QRS information. Then the results were compared against the annotations provided by the cardiologists on the MITDB database. Further, in \cite{td6, td7, td8} authors extracted the ECG features and classified various ECG anomalies based on the 12 Lead ECG signal. The advantage of this algorithm is forward search processing; it lowers the memory requirement in real-time processing. Similarly, in \cite{td9, td10} authors detected the ECG features and classification scheme by which various diseases were possible to detect. The results of real time system were not reported in the work.

In conclusion, temporal features such as RR, PR, QT interval, ST deviations were utilized to provide the diagnostic results. Various filtering methods such as differentiator, moving average filter, low pass, high pass etc. have been utilized to obtain ECG signals' feature points. 
The time domain methods are of prime importance as the cardiologist detects the heart's anomalies using 12- Lead ECG signal in the time domain that provides synergistic efforts between doctors and engineers towards automated detection systems. 

\begin{table}
\centering
\fontsize{6}{8}\selectfont
\caption{Time domain processing of ECG signals}
\vspace{2 mm}
\label{Timedomain}
\begin{tabular}{c c c c c c }
	\hline
	
	\rowcolor{lightgray} \textbf{Reference} & \textbf{ Research Orientation} & \textbf{Database}  & \textbf{Classifiers} &  \textbf{Performance}  \\
	\hline
	\ \cite{Tompkins2} & Real Time &  MITDB & Adaptive & FDR\%= .675 \\
	\ &  QRS Detection & &  Thresholding \\
	\hline
	\ \cite{td1} & Real Time  &  MITDB &  Combined  & Se\%= 99.69,\\
	\  & QRS Detection & &  Adaptive & Sp\%= 99.66     \\
	\ & && Thresholding & (Algorithm1)   \\
	\ &&&& Se\%= 99.74 \\
	 	\ &&&& Sp\%= 99.65 \\
	 \ &&&&	(Algorithm2)\\
	\hline

	\ \cite{td2} &  QRS Complex  &  MITDB & Difference  & FDR \%= 0.19  \\
	\ & Detection & & Operation   \\
	\ &&& Method \\
	\hline
	\ \cite{td3} &  R peak detection &  PTB & ECG double  & Se \%= 99.8  \\
	 \ & Algorithm & &  difference  and  \\
	 \ & & & RR interval  \\
	 \ &&& processing\\
	
	\hline 
	\ \cite{td11} & ECG feature &  QTDB, & Time Domain   & - \\
	 \ &  Extraction &  PTB & Gradient based &\\
	 \ &&& classification \\
	 
	\hline
	\ \cite{td4, td5} &  P and T wave  & MITDB  & Based on two & Se \%= 98.05,   \\
	\ & Detection && moving average   & PPV\%=97.11     \\
	\ & & & filters Dynamic  & (P wave) \\
	\ & & & event related  & Se \%= 99.86,  \\
	\ &&& thresholding &  PPV\%=99.65 \\
	\ &&&& (T wave) \\
	\hline
	\ \cite{td6, td7, td8} & Disease diagnostic   & INCARTDB    & Decision Logics  & FDR \%=0.69,  \\
	\ & algorithm based &  PTB and & and Empirical  & 0.69, 0.34      \\ 
	\ &  on forward search & QTDB & Classifiers & \& 1.72 for BBB,   \\
	 \ &&&& Hypertrophy, \\
	 \ &&&& Arrhythmia \\
	  \ &&&& MI \\
	
	\hline
	\ \cite{td9, td10} & Disease diagnostic  & MITDB & Decision Logics  & FDR \%=1.289\%, \\
	\ & algorithm &  & and Empirical  & PPV \%= 99.293,\\
	\ &&& Classifiers  &  Se\%=99.492 \\
	\ &&&&  (QRS Detection)\\
	\hline
	
\end{tabular}
\end{table}

\subsubsection{Frequency Domain}
Research onthe frequency domain processing of ECG signals started around the 1970s.
During the initial years, the researchers were focusing on obtaining the frequency band amplitude information related to ECG signals such as the band of QRS complex, P wave and T wave \cite{fd1, fd2} \textit{etc}. 

The features required for diagnosis are temporal and statistical information in nature. Hence, ECG signal processing based on only spectral information has lower efficiency than the time domain and time-frequency methods. The frequency domain methods are shown in Table \ref{freqdomain}. As can be seen in \cite{fd4, fd7, fd8, fd9, fd10} research, results were not available in terms of standard performance matrices, which makes it quite difficult for comparison. In \cite{fd5}, the ECG signals were classified in different categories namely ventricular fibrillation and flutter (VF), Artifacts (A), series of complexes of aberrent morphology (CAM) and one unknown category by utilizing the spectral features of the ECG signal. However, the ECG data used for the purpose was consists of only 55 ECG signals from an unknown database. In \cite{fd7-0} authors detected the QRS, P and T waves of the ECG signals by application of discrete fourier transform on the ECG signals, acquired locally and from MITDB. The delineated components were evaluated visually and by computing the Normalized Mean Square Error between the original signal and recreated signal. In \cite{fd7-01} authors detected the duration of ventricular fibrilation in swines and humans using the frequency domain features of ECG signal. In \cite{fd7-1} determination of counter shock (Medical Procedure) success  was done by obtaining the fourier transform of ECG signal and it was concluded that the median frequency, dominant frequency and amplitude of the signal can predict the success or failure of the procedure.

In \cite{fd11}, authors determined the power line interference, a common source of noise in the ECG signal and leads to imprecise measurements of the ECG wave durations and amplitudes. In \cite{fd12} authors determined the heart rate variability using spectral analysis. However the results are not validated on any standard databases. In \cite{fd13}, authors detected  anomalies of ECG signal by utilizing the frequency domain features and the results are compared with the temporal methods.  

Frequency domain features are essential to obtain the complete analysis for the signal. However, the disease diagnosis require the utilization of temporal features as most of the features for cardiac anomalies were developed by cardiologists and physicians who are accustomed to time domain signals.

\begin {table}
\fontsize{6}{8}\selectfont
\caption{Frequency Domain Processing of ECG signals}
\vspace{2 mm}
\label{freqdomain}
\rotatebox{90}{
\begin{tabular}{c c c c c c}
	\hline
	
	\rowcolor{lightgray} \textbf{Reference} & \textbf{Research Orientation} & \textbf{Database}  & \textbf{Classifiers} &  \textbf{Performance}  \\
	\hline
	\ \cite{fd1}  & Determination of bandwidth of ECG  & 10 normal ECGs  & -  & Waveform- Amplitude 0-200 Hz\\
	\   & by applying Fourier Transform & Database- Unknown & & Waveform duration- 0-60Hz \\
	\hline
	\ \cite{fd2} & Determination of maximum frequency   &  Healthy subjects' ECG &  &$F_{max}$ P wave = 55Hz,
	
	\\
	\   & components in P, QRS and T waves   & Database- Unknown & & QRS-complex = 65 Hz,  \\
	\ & & & & T wave-25 Hz\\
	\hline
	\ \cite{fd4} & Power spectrum Analysis of HRV & 18 subjects ECG recordings & \\
	\   &  for Sudden Cardiac Death & Database-Unknown & -  \\
	\hline
	\ \cite{fd5} & Ventricular Arrhythmia (VF), Artifacts (A)  & 55 ECG recordings & various spectral features &  VF Sp= 70\% Se= 100 \% \\
	\ & and Series of complexes of  & Database- Unknown &  & CAM Sp=  100\% Se=86 \% \\
	\ & aberrent morphology (CAM) &&  & A Sp= 100\% Se=92\% \\
	\hline
	\ \cite{fd7} & Frequency Analysis of ECG & Bipolar Leads & Maximum Entropy Method &  \\
	\ &   for Tachycardia Detection  & Unknown Database & based on Autoregressive model & -  \\
	\hline
	\ \cite{fd7-0} & P, QRS and T wave & locally acquired ECG signals & Discrete  Fourier & -\\ 
	\ & detection of ECG signal & and MITDB & Transform & \\
	\hline
	\ \cite{fd7-01} & Ventricular Fibrillation duration & Unknown & Frequency domain & Estimation = 82.9 \% \\
	\  & Detection in swines and &  & Features such as & \\
	\ & humans  & & median frequency & \\ 
	\hline 
	\ \cite{fd7-1} & Prediction of Counter & 26 patient's ECG & Dominant frequency,&  Se\%= 100\\
	\ & shock therapy's success by & Unknown Databse & median frequency and & - \\
	\ &  fourier transform of the signal &  & amplitude & - \\
	\hline
	\ \cite{fd8} & ECG spectrum feature for & MITDB & Frequency domain &  Approximate  pobablity of error  \\
	\ & classification of arrhythmia & & features &  0.0\%\\
	
	\hline
	\ \cite{fd9} & Method for removing artifacts  & Real Time ECG & Convolutive Indepent  &  -\\
	\ & & obtained from Smartex system & Component Analysis model &   \\ 
	\hline
	\ \cite{fd10} & Frequency domain Analysis  & MITDB and & Auto Associative & -  \\
	\ & &  SVDB & Neural Networks \\ 
	\hline 
	\ \cite{fd11} & Detection of Power & 12000 ECG Signals & Power Spectral   & FP Rate= 0.1\% \\
	\ & line interference &  of 10 sec duration & Density & \\
	\hline
	\ \cite{fd12} & HRV and Breath to Breath & 5 female and 6 male subjects  & Auto Regressive Moving Average &  peak amplitude (0.15-0.40 Hz) band $>$
	\\
	\ &  Interval Analysis & Unknown Database & Model for Spectral Analysis & frequency (0.04-0.15 Hz) band \\
	
	\hline 
	
	\ \cite{fd13} & Disease Diagnosis in &  MITDB & Frequency based &  80\%  lesser ECG diagnosis \\
	\ &  Frequency domain & & Neural Networks &   time at 5\% Accuracy lose \\ 
	\ & & & &  compared to temporal methods\\
	\hline
\end{tabular}
}
\end{table}

\subsubsection{Time-Frequency Domain }
Time-frequency domain methods utilize the time as well as frequency space for obtaining the features simultaneously. The most popular technique in this domain is Wavelet Transform (WT) \cite{WTreview}. The advantage of using WT over Fourier transform is that Fourier Transform of a signal provides the information regarding the frequency and its magnitude however, it cannot provide frequency information for a localized signal in time.
To overcome the poor time resolution of the Fourier transform, Short Time Fourier Transform (STFT) has been developed. It provides the time-frequency representation of the signal. In the Wavelet Transform signal's different frequency components are analyzed at different time resolutions also known as multiresolution analysis \cite{WTbook}. The multiresolution analysis capability of WT makes it suitable for biomedical signal processing schemes \cite{WTreview}. Denoising technique for an ECG signal using new wavelet- and wavelet packet-based schemes were discussed with simulated noises in \cite{tfd0}. 
Various WT techniques such as Cross Wavelet Transform (XWT) \cite{tfd8}, Continuous Wavelet Transform (CWT) \cite{tfd1} and Discrete Wavelet Transform (DWT) \cite{tfd3, tfd4, tfd5, tfd6,tfd7, tfd8, tfd9, tfd12, tfd13, tfd14} are employed in literature \cite{tfd1, tfd2, tfd3, tfd4, tfd5, tfd6, tfd7, tfd8, tfd9, tfd10, tfd11, tfd12, tfd13, tfd14}.   
Various researches for WT and other time-frequency schemes are given that gives a comparison in Table \ref{Timefreqdomain}. In \cite{tfd1}, authors report detection of the arrhythmias in the ECG signals of pigs. WT was utilized for the compression of ECG signal in \cite{tfd2, tfd3, tfd4}. The main reason for utilizing WT for the compression is because it transforms the signal's energy into fewer transform coefficients. So, most of the transform coefficients with lower energy can be discarded \cite{tfd4-2}. It is also an efficient and flexible scheme for compression \cite{tfd4-3}. Authors in \cite{tfd5, tfd6, tfd9, tfd11} detected various features such as R peaks, QRS complex using WT. In \cite{tfd7, tfd7-2, tfd8, tfd10, tfd12, tfd13, tfd14} authors classified the ECG signals by utilizing the wavelet coefficients as features. In \cite{tfd8-2} two algorithm were presented based on Wavelet threshold based TDL and TDR algorithms for real-time ECG signal compression. The authors achieved the correct diagnosis (CD) values upto 100\%  for various compression ratios.
 The Wavelet Transform too has certain disadvantages as it becomes computationally intensive for finer resolution. Hence, Discrete Wavelet Transform (DWT) offers fast computations due to the discretization of wavelets as the minimum energy transform coefficients are discarded at the cost of efficiency of the system \cite{WTbook}.

\begin {table}
\centering
\caption{Time-Frequency domain features of ECG signals}
\fontsize{6}{8}\selectfont
\vspace{2 mm}
\label{Timefreqdomain}
\rotatebox{90}{
\begin{tabular}{c c c c c c }
	\hline
	
	\rowcolor{lightgray} \textbf{Reference} & \textbf{Research Orientation} & \textbf{Database}  & \textbf{Classifiers} &  \textbf{Performance}  \\
	\hline
	\ \cite{tfd1}  &  Arrhythmia Detection  & Pigs ECG   &  - &- \\
	\ & in ECG utilizing WT & signals & \\
	
	\hline
	\ \cite{tfd2}  &  Compression of ECG & MITDB  & -  & \% Root Mean Square\\
	\ & using WT &  &  &  Difference=1.18 (1 dataset)  \\
	
	\hline
	\ \cite{tfd3} &   Compression of ECG & MITDB & - & Bit Error Rate= $10^{-15}$ (3 datasets) \\
	\ & using WT &  &  & \% Reduction Transmission Time= 72.7  \\
	
	\hline
	\ \cite{tfd4} & Compression of ECG & MITDB & - &  \% Root Mean Square\\
	\ & using WT &  &  &  Difference=1.08 (1 dataset)  \\
	
	\hline
	\ \cite{tfd5} & Single Lead ECG  & EDB,  & Multiscale Approach & Se\%= 99.66, PPV\%= 99.56 (ST-T, CSE)\\
	\ & delineation system & CSE, MITDB & & Se\% and PPV\% = over 99.8\\
	\hline
	
	\ \cite{tfd6} & ECG QRS Detection & MITDB & Two wavelet filters (D4 and D6) & Se\%=$99.18 \pm 2.75$   \\
	\ & using Multiresolution WT & & for QRS detection &   PPV\%= $98.00 \pm 4.45$ \\
	
	\hline
	\ \cite{tfd7}& Personal Identity Verification  & QTDB & Euclidean distance measure & False Acceptance Rate \\
	\ & with WT of ECG Signal & & for verification & = 0.83\%(N) and 12.50\% (Ab)\\
	\ & & & &	False Rejection Rate \\
	\ & & & &		= 0.86\%(N) \&  5.11\%(Ab) \\
	\ & & & & (N=Normal Ab= Abnormal) \\
	\hline
	 \cite{tfd7-2} & Denoising of ECG  & MITDB & Adaptive Thresholding & Se= 99.71\% \\ 
	 \ & and QRS Detection &  & on DWT & Sp=99.72\% \\
	 \ & & & & DER = 0.52\% \\
	\hline
	
	\ \cite{tfd8} & Cross WT for classification & PTB & Parameter obtained from  & Acc\%=97.6, Se\%= 97.3\\
	\ & and analysis of ECG signals &  & XWT wavelet cross spectrum   & Sp\%= 98.8 \\
	\ & & & and	wavelet coherence & \\
	
	\hline
	\ \cite{tfd8-2} & Wavelet based ECG & MITDB, & Wavelet threshold & CD=100\% \\
	\ &  Signal Compression  & SVDB  & & based & \\  
	\hline
	
	\ \cite{tfd9} & Low-Complexity ECG Feature & QTDB, PTB, & DWT with Haar & $2.423N + 214 (+)\  and \ 1.093N + 12 (\times) $\\ 
	\ &  Extraction & a Non commercial  & and mother wavelet  &  for $(N \le 861)$   \\
	\ & & Database & functions &  $ 2.553N + 102(+) and 1.093N + 10 (\times)$ \\
	\  & & & & $ for (N >861)$\\
	\ & & & & (N= number of input samples)\\
	\hline
	
	\ \cite{tfd10} & Wavelet based ECG detector & MITDB & Soft-threshold algorithm & Detection Error-Rate\%=  0.196 \\
	
	\hline
	
	\ \cite{tfd11} &  ECG Recording and & MITDB  &  R peak detection & Se\%= 99.72, PPV\%=99.49 \\
	\ & R-Peak Detection Based on WT & &  with a FIR filter & Data Reduction =13.68 $\times$ \\
	
	\hline
	\ \cite{tfd12} & ECG-Based Biometric  & CEBSDB, WECG,  &  Multiresolution
	convolutional  & Average Identification Rate \% \\
	\ & Human Identification  & NSRDB, STDB, & neural network &  = 93.5 \\
	\ & & AFDB,  VFDB & & \\
	\ & & FANTASIA, MITDB & & \\
	\hline	
	\ \cite{tfd13} & ECG Abnormality  & MITDB  and   & Support Vector Machine & Maximum Accuracy= \\
	\ & Detection & real time data &  & 96\% \\
	
	\hline
	\ \cite{tfd14} & Time- Frequency domain & St. Petersburg 
	& Random tree and J48  & Acc\%= 99.93\\
	\ & coronary artery disease & and FANTASIA & decision
	tree for time  &\\
	\ & detection  & & frquency domain features  & \\
	
	\hline
\end{tabular}
}
\end{table}

\subsection{Classifiers}
The Classification considers the extracted feature points from the previous stage and classifies the signal into different categories by using empirical, adaptive and constant threshold-based , machine learning-based and Deep Neural Networks based classifiers.
The conventional empirical classifiers are generally based on the medical observations for the particular field.  Thresholding based or decision logic-based approach is based on defined logical rules, \textit{e.g.} R-R interval, ST interval \textit{etc.} 

The machine learning-based approaches based on multivariate statistical pattern recognition has a widespread utilization in biomedical signal processing. These methods utilize correlation analysis, regression techniques and template matching to identify abnormal patterns or a particular class of signals \cite{mnreview,mnreview2}. However, as these statistical methods move towards greater accuracy, it also has a higher system complexity. Most recent techniques are deep neural networks also known as an artificial neural network (ANN) consists of multiple hidden layers between the input and output layers \cite{mnreview3}. Each layer consists of neurons with different weights and biases. The neurons can pass the information to other neurons in other layers. The backpropagation technique provides feedback and updates the weight associated with neurons offering supervised and unsupervised learning.  The deep learning technique offers more accuracy to the system at the cost of increased system complexity. 
Recent developments in deep neural networks is widespread, with the latest techniques discussed in \cite{mnreview3} are Recurrent Neural Networks (RNN), Convolution Neural Networks (CNN) and other generative models  such as Autoencoders and 
Generative Adversarial Network (GAN). In the following subsection, various machine learning and neural networks based classifiers for ECG signal processing are discussed.

\subsubsection{Machine Learning and Deep Learning Technique based Classifiers}
This section categorizes (shown in Table \ref{machine}) various researches based on machine learning and deep learning techniques for classification. 
In \cite{mn1}, authors provided the customized ECG classifier with patient-specific data based on an unsupervised learning technique. The method's limitation was that it required to develop a local classifier for each patient with patient-specific data. 
In \cite{mn1-1}, authors detected the QRS complexes of 12 Lead ECG signal available in CSE dataset -3  with a supervised learning of ANN. The back propagation algorithm has been used to train the system. 
Authors in \cite{mn2}, utilized the ANN for arrhythmia classification, ischemia detection, and recognition of chronic myocardial diseases. It used both static and a recurrent ANN with preprocessing and postprocessing that defined the dimensions of input features for neural networks.
Authors in  \cite{mn2-2} utilized an unsupervised learning clustering scheme for the classification using Hermite functions based features of QRS complexes. The limitation was that it did not provide signal quality information in the input vector's self-organizing maps.
Authors in \cite{mn3}, used a beat recognition and classifier based on a supervised learning scheme that utilized fuzzy hybrid neural network and higher-order statistics features as inputs.
In \cite{mn4}, authors utilized Hermite basis function expansion of the QRS complexes of ECG waveforms and modified Takagi-Sugeno-Kang neuro-fuzzy network for heartbeat recognition and classification based on a supervised learning scheme.
In \cite{mn5}, authors utilized a popular supervised machine learning approach known as Support Vector Machine (SVM) for the recognition purpose. The input features in the method were obtained by two methods namely higher-order statistics (HOS) and Hermite characterization of the QRS complex. 
In \cite{mn6}, authors classified the ECG data into three categories, namely normal beat, ventricular ectopic beat(VEB), supraventricular ectopic beat (SVEB). The classification was based on a statistical classifier model utilizing a supervised learning scheme. The limitation was the heartbeat fiducial points were manually annotated.
In \cite{mn7}, authors provided supervised learning based on a decision tree based classifier algorithm to be implemented on a personal digital assistant (PDA). But, the algorithm was not implemented in a real time environment as the PDA was meant only for a system demonstration.
In \cite{mn7-1}, authors used features such as ST segment area, R-S interval, ST-slope, R-T interval, QRS area, Q-T interval, R-wave amplitude, heart beat rate and four statistical features QRS energy, mean of the power spectral density, auto-correlation coefficient,and signal histogram are applied to signal stage and two stage feed forward neural networks for the anomalies detection.
In \cite{mn8}, authors used supervised learning that required block based neural networks as classifiers. It utilizes Hermite coefficients and R-R intervals as input features to classify Supraventricular Ectopic beats and ventricular ectopic beats.
Authors in \cite{mn9} utilized the supervised particle swarm optimization (PSO) with the support vector machine classifier on the automatically detected features.
In \cite{mn10}, authors compressed the ECG signal using local extreme extraction,
adaptive hysteretic filtering and Lempel–Ziv–Welch (LZW) coding. The reconstructed waveform was verified with frequent normal and pathological types of cardiac beats using a multilayer perceptron neural network trained with original
cardiac patterns and tested with reconstructed ones.
In \cite{mn11}, authors did the screen apnea screening using the time domain and frequency domain features. It used two approaches, namely K-nearest neighbor (clustering or unsupervised technique) and Neural networks that offer the supervised learning scheme. The limitation of the method was that it was unable to detect isolated apneas and other physiological and pathological events during
sleep, such as cyclic alternating pattern and periodic leg movements that could affect the classifier's efficacy.
In \cite{mn12} authors used local fractal dimension (LFD) of neighboring
sample points of ECG signal segments are used as the features. For estimating the LFD, two different methods namely power spectral density based fractal
dimension estimator (PSDFE) and variance based fractal dimension
estimator (VFE) are used.

In \cite{mn13}, authors obtained the ECG features set  with sequential
forward floating search algorithm based on linear discriminants. The most suitable subset was again evaluated with a multilayer perceptron Neural network as a supervised learning scheme. The selection of suitable features led to complexity reduction for the system. In \cite{mn14}, mobile-cloud based electrocardiograph monitoring scheme was compared with the mobile-based systems. The system utilized the supervised ANN for classification. In \cite{mn14-1}, Particle Swarm Optimization (PSO) based wavelets are applied to the Support vector machine for categorizing various ECG signals. 
In \cite{mn14-2}, three neural network classifiers  Back Propagation Network (BPN), Feed Forward Network (FFN) and Multilayered Perceptron (MLP) are utilized for ECG anomlies detection.
In \cite{mn15}, authors classified the ECG signals into Supraventricular Ectopic beats and ventricular ectopic beats with the supervised  1-D Convolutional Neural networks and the patient-specific data. The method's limitation was that the dedicated CNN was trained for an individual patient and often posed a challenge in the Real-time environment. In \cite{mn16}, authors designed a personalized heartbeat classification model for long ECG Signals and implemented the system with parallel general regression neural network (GRNN). Further to this, they implemented an online-learning module into the parallel GRNN for the real-time personalized automatic classification on the Holter ECG data. In \cite{mn17}, authors implemented various machine learning schemes such as end-to-end CNN, KNN, linear SVM, Gaussian kernel SVM, and Multilayer perceptron classifiers for categorizing the paroxysmal atrial fibrillation cases and concluded that integration of convolution neural network as a feature extractor with other conventional neural network-based classification methods provided better results. 
In \cite{mn17-2}, bidirectional Long-short term memory networks based wavelet sequences called DBLSTM are used for categorizing various arrhythmic signals such as Normal Sinus Rhythm (NSR), Ventricular Premature Contraction (VPC), Paced Beat (PB), Left Bundle Branch Block (LBBB), and Right Bundle Branch Block (RBBB) available in the MITDB Database. However, the work did not validate the results under the noisy input conditions. In \cite{mn17-3},  authors  used K-Nearest Neighbour (KNN) and Principal Component Analysis (PCA)  classifier techniques with auto regressive modelling schemes to categorize Atrial Tachycardia, Premature Atrial Contractions and Sinus Arrhythmia.

In \cite{mn18}, authors converted the time domain ECG signals into time-frequency domain spectrogram by utilizing Short-time Fourier transform and the spectrogram was utilized as the input to 2-D and 1-D  convolution neural network for the arrhythmia classification on the MITDB database. In \cite{mn20}, authors integrated a long short-term memory based auto-encoder (LTSM-AE) network for features learning with a support vector machine for electrocardiogram (ECG) arrhythmias classification and followed a supervised learning scheme. In \cite{mn21}, authors developed a system to categorize the severity stages of Myocardial Infarction condition and utilized an attention-based recurrent neural network for automated diagnosis of the three MI severity stages by processing the 12 Lead ECG data. The work's limitation was that it addressed only the classification of MI severity stages, not the diagnosis of the conditions.

As can be seen in the works mentioned above, most of the methods utilize the databases as the input signal and are processed as a software-based approach. Most systems offered higher efficiency and provided less error and higher accuracy with increment in computational load. The methods generally adopted the supervised learning schemes for the classification on the databases that added constraint on the system level implementation because for the real time data, the system needs to be trained on real time ECG data. 

\newpage

\fontsize{6}{8}\selectfont
\begin{longtable}{c c c c c}
    
	\caption[]{Machine Learning and Neural Network for ECG Signal Processing} \label{machine} \\
	
\hline
 \ \textbf{Ref.} & \textbf{Research } & \textbf{Database}  & \textbf{Classifiers} &  \textbf{Performance}  \\
 \ & \textbf{Orientation} \\
	\hline
	\cite{mn1} & ANN Based & MITDB  & Self organizing   & CR \%= 94.0,   \\
	\ &  Classification  &  &  maps ,learning  & Se\%= 82.6,   \\
	\ & & & vector quantization,  & PPV\%= 77.0\\
	\ &  & & mixture-of-experts  & Sp\% = 97.1\\
	\ &&& (Patient Adaption)\\
	\ &&& (Unsupervised Learning) \\

	\hline
	\cite{mn1-1} & QRS Complex detection & CSE & Artificial Neural& Se \%= 99.11\\
	\ & of 12 Lead ECG signal && Network \\
	
	\hline
	
	\cite{mn2} & Arrhythmia  & MITDB,   & Static and Recurrent & -\\
	\ & classification,   & CSE, & Artificial Neural Networks & \\
	\ & ischemia detection,  & EDB \\
	\ & \& recognition of  \\
	\ &  chronic myocardial \\
	\ & diseases \\ 
	\hline
	\cite{mn2-2} & Clustering of various & MITDB & Unsupervised self  & CER\%=1.5 \\
	\ &  ECG complexes using  & & organized Neural &  \\
	\ & Hermite Basis && networks  \\
	\ & functions features\\
	
	\hline
	
	\cite{mn3} & Recognition and & MITDB & C-means and 
	& ER \%= 2.55 \\
	\ &  classification of & & Gustafson-Kessel & (Training)\\
	\ & different type && 	algorithms with & ER\%= 3.94\\
	\ & of heart beats &&  fuzzy hybrid & (Testing)\\
	\ &&& neural network \\
	\hline
	\cite{mn4} & On-Line Arrhythmic & MITDB & Hermite coefficients & EOR\%=96  \\
	
	\ &  Heart Beat  & &  features to  & (Testing) \\
	 \ & Recognition Using & & Fuzzy Neural \\ 
	 \  & Hermite Polynomials && Networks  \\
	\hline
	
	\cite{mn5} & Higher Order & MITDB & Support Vector  & ER\%= 6.28 (HOS)\\
	\ &  Statistics (HOS) and  & & Machine & (Testing)\\
	
	\ & Hermite Preprocessing &&&  ER \% = 5.43 (HP)  \\
	\ & (HP) Features for &&& (Testing) \\
	\ & Automatic classification \\
	\hline
	
	\cite{mn6} & Automatic  & MITDB & Classifier based  & SVEB Se\%= 75.9,\\
	\  & Classification  of   & & on  Linear  & , PPV\%= 38.5   \\
	\ & Heart Beats using   & & Discriminants & FP \%= 4.7 \\
	\ & morphology and & & &    VEB, Se\%= 77.7,\\
	\ & heart rate  & & &  PPV\%=81.9   \\
	\ & interval features &&& FP \%= 1.2\\
	\hline
	
	\cite{mn7} & Classification of ECG  & MITDB & Decision Tree or & Acc\%= 100  \\
	\ & on Personal Digital & & J48 Classifier & (Critical Arrhythmias)\\
	\  &  Assistant &&& Acc\%=97.95  \\
	\ &&&& (Arrhythmias) \\
	\ & & & &   Acc\%= 95 \\
	\ &&&& ( Arrhythmias before \\
	\ &&&& Critical cases) \\
	
	\hline 
	\cite{mn7-1} & Time, frequency and   & MITDB &  Two-stage feed forward  & Highest Recognition \\ 
	\ & statistical features utilization   & & neural network  & Rate= 93\%\\
	\  & for Anomalies Detection &  &&\\
	\hline

	\cite{mn8} & Personalized ECG  & MITDB & Block Based&  SVEB Detection\\
	
	\ & classification on  & &  Neural   &  Acc\%= 96.6 \\
	\ & Hermite Coefficients  & & Networks &   VEB Detection\\
	\ & and R-R interval  & & &  Acc\%= 98.1\\
	\ & features \\
	\hline
	
	\cite{mn9} & ECG classification & MITDB & Support Vector& Detection \\
	\ &  based  on & & Machine with   & Acc\%=  89.72\\
	\ & automatically && Particle Swarm \\
	\ & detected features && Optimization \\
	\hline
     
	\cite{mn10} & ECG compression  & MITDB & Principal Component & Classification\\
	\ &  and Classifiaction & & Analysis &  Rate\%=90 \\
	
	\hline
     \newpage
     \hline
	\cite{mn11}  & Sleep Apnea  & APNEA- & K-nearest neighbor  &  KNN-Acc\%=88,   \\
	\ & screening using  & ECG &  (KNN) and Neural & Se\%= 85,  \\
	\ &  temporal and  & &  networks (NN)  & Sp\%=90 \\
	\ & spectral features & & supervised learning & NN-Acc\%= 88, \\
	\ &&&& Se\%= to 89, \\
	\ &&&& Sp\%=86  \\
	
	\hline
		\cite{mn12} & Local Fractal   & MITDB & PSDFE  & PSDFE  Se\%= 95.6 \\
		 \ & Dimension based   &  &   and VFE &  Sp\%=99.4   \\
		   \ & Arrhythmia Classification & & & VFE  Se\%= 92.8\\
		   \ &&&& Sp\%=98.8\\  
	\hline
	\cite{mn13} & ECG Arrhythmia  & MITDB & Linear Discriminant(LD) & LD (SFFS) \\
	\ & classification by   &  & MLP Neural Networks & Acc\%=84.6 \\
	\ & Temporal, Statistical,  & & with Sequential Forward & MLP(SFFS) \\
	\ & Morphological features & & Floating Search (SFFS) & Acc\%= 89\\
	\ & selection \\
	
	\hline
	
	\cite{mn14} & Personalized ECG  & MITDB & Artificial Neural & Classification \\ 
	\ & Telemonitoring   & & Networks with & Acc\% =98.82   \\
	\ &&& 30 Neurons&  (Training) \\
	\ &&&& Classification \\
	\ &&&&  Acc\%= 63.92 \\
	\ &&&& (Testing) \\
	\hline
	\ \cite{mn14-1} &  PSO Wavelet based features &  MITDB & Support & Se\%=91.75 \\
	\ & utilization for signal & & Vector Machine & Sp\%=96.14 \\
	\ & classification & & &  PPV\%=74.26 \\
	
	\hline
	 \ \cite{mn14-2} & DWT based & MITDB & BPN, FFN and MLP & Acc\%= 100\\
	 \ & morphological features  & based ANN & \\
	 \ & utilization for  & & & \\
	 \ & abnormality detection & & & \\
	    
	     \hline
	\cite{mn15} & Real-Time  & MITDB & 1-D Convolution & SVEB \\ 
	\ & Patient Specific&  & Neural Networks& Acc\%= 99\\
	\ &  ECG Classification &  & & VEB \\
	\ & & & & Acc\%=97.6\\
	
	\hline
	\cite{mn16} & Personalized Heart  & MITDB & Parallel general 
 &  Classification \\
	\ & Beat classification && 	regression neural & Acc\%= 95 \\  
	\ &  on long term ECG && network & (Database) \\
	\ &&&& Acc\%= 88\\
	\ &&&&  (Real Time)\\
	\hline

	\cite{mn17} & Atrial Fibrillation  & AFPDB & Convolution Neural & Precision\%\\
	\ & detection on  & & Networks with   & =  90.65 \\
	 \ & automatically detected && Deep Learning \\
	 \ & features  \\
	
	\hline
	\cite{mn17-2} & Arrhythmia  & MITDB & Wavelet sequence based & Recognition performance \\
	\ & Classification && on deep bidirectional & =99.39\% \\
	\ &&& LSTM network & \\
	\hline
	\cite{mn17-3} & ECG signal   & Unknown & PCA and KNN with & Acc\%= \\
	\  & classification & & auto regressive modelling &  upto 100 \\ 
	\hline
	\cite{mn18} & Arrhythmia & MITDB & Deep 1-D and 2-D  &  Acc\%=99  \\
	\ &  Classification  & & Convolution Neural &  (2-D CNN)\\
	\ &&&  Networks(CNN) &  Acc\%=90.93 \\
	\ &&&&  (1-D CNN)\\
	\hline
	\cite{mn20} & Arrhythmia  & MITDB & Support Vector & Acc\%= 99.74,  \\
	\ & Classification   & &  Machine & Sp\%= 99.84, \\
	 \ & with long  &&& Se\%= 99.35 \\
	 \ & short-term  \\
	\ & memory based \\
	\ & auto-encoder  network \\
	\hline
	
	\cite{mn21} & Myocardial Infarction  & PTB & Multi-lead  & Detection \\
	\ & severity stages  &  & Diagnostic Attention & Acc\%= 97.79\\
	\ & classification &&  based Attentional \\
	\ &&&  Recurrent Neural \\
	\ &&& Network \\
	\hline 
\end{longtable}
\footnotesize{$^{SVEB}$ Supraventricular Ectopic Beat, $^{VEB}$ Ventricular Ectopic Beat}\\

\section{ Discussion and Conclusion}
This review report provides insight into the invention of ECG, global acceptance of ECG, the evolution of leads and transformation of the system from the huge String galvanometer to portable monitors in hospital settings.

Additionally, it offers a view on the earlier ECG signal processing schemes to the most recent ones. The signal processing can be mainly categorized into four steps starting from acquiring the ECG signal, preprocessing of ECG signal, Feature Extraction in different domains and classification schemes. 

In the course of this review, the authors' conclusions are listed below:

1. The machine learning, deep learning techniques and wavelet transforms are designed on the software processing domains that pose a constraint for the real time systems for battery requirements.

2. The researches utilized the databases available on Physionet for data acquisition that are old dated. For example, the widely used MIT-BIH arrhythmia database was recorded during 1975-1980. Over the last 40 years span, various standard definitions for the diseases changed according to various standards viz. American Heart Assocition(AHA), European Society of Cardiology (ESC) \textit{etc}. The use of the annotations provided in the database for comparing the results with the automatic means therefore remains slightly questionable. The need of inclusion of most recent databases or improvements of annotations with the recent standards may lead to more accuracy.

3. Lead uniformity is missing with the automatic methods discussed. For example, MIT-BIH arrhythmia database provides the MLII lead and V5 leads and other  databases such as PTB provides the 15 Lead data. Some of the methods are tested on the 12 Lead data and others are on 2 Lead data. 

4. The literature does not utilize patients information (age, gender, previous medical history \textit{etc}.) provided with the dataset that can add additional variables for the classifiers and can lead to more efficient systems, according to European society of cardiology.

5. Utilization of supervised learning classifiers are suitable to obtain results on the databases. For system implementation, the data needs to be trained and tested based on the actual data and that is a challenge in the case of supervised learning schemes. 
 
 6.  The demographics conditions of the countries or regions may vary, hence the algorithm developed to diagnose for specific region of people must be adaptable to these conditions, that can be validated particularly with the locally obtained database. The standard/existing databases does not address this.
   
\section*{Acknowledgement}
The authors would like to thank Dhirubhai Ambani Institute of Information and Communication Technology for the research support. This work is supported by  Science and Engineering Research Board (SERB) (Grant Number:DST SERB Grant CRG/2019/004747) under Department of Science and Technology (DST), Government of India.

\end{document}